\documentclass[reqno, 11pt]{amsproc}

\newtheorem{result}{Result}[section]
\newtheorem{lemma}{Lemma}[section]
\usepackage{amsmath}
\usepackage{subeqnarray}
\usepackage{graphicx}
\usepackage{natbib}
\usepackage[width=1.2\textwidth]{caption}

\usepackage[a4paper, total={6.22in, 9in}]{geometry}
\usepackage{fancyhdr}
\usepackage[font=footnotesize,labelfont=bf]{caption}
\usepackage{hyperref}

\providecommand\bnabla{\boldsymbol{\nabla}}
\providecommand\bcdot{\boldsymbol{\cdot}}

\title{\raggedright Pumping and Steady Streaming driven by Two-Frequency Oscillations of a Cylinder}

\author{
\begin{flushleft}
$\text{Hyun S. Lee}^{1}, \text{William D. Ristenpart}^{2}$\and$\text{ Robert D. Guy}^{1}$\\
\tiny
\text{}\\
1 Department of Mathematics, University of California, Davis, CA 95616\\
2 Department of Chemical Engineering, University of California, Davis, 95616\\
\text{}\\
Corresponding Author: Robert D. Guy, \href{mailto:rdguy@ucdavis.edu}{rdguy@ucdavis.edu}
\end{flushleft}
}
\normalsize
\begin{document}
\begin{abstract}
The classical problem of steady streaming induced by an oscillating object has been studied extensively, but prior work has focused almost exclusively on single-frequency oscillations, which result in symmetric, quadrupole-like flows. 
Here we demonstrate that dual-frequency oscillations induce asymmetric steady streaming with a non-zero net flux in a direction determined by the polarity of the oscillation \--- the oscillator serves as a pump. 
We use numerical simulations and asymptotic analysis at small amplitude to examine 2D steady streaming around a cylinder, first focusing on frequency ratio two. 
The computational experiments show asymmetrical streaming and pumping, i.e., net flux downstream. 
It is well known from asymptotic analysis that steady streaming is second order in amplitude, and we show pumping occurs at third order.
We then extend the analysis to general frequency ratios, where we give necessary conditions for pumping and predict the order in amplitude at which pumping occurs. 
Finally, we corroborate the theoretical results with computational simulations for different frequency ratios, and we discuss the implications for using dual-mode vibrations to pump fluids in lab-on-a-chip and other applications.
\end{abstract}
\maketitle

\section{Introduction}
Steady streaming induced by an oscillating object is a classical phenomenon that has been studied analytically and experimentally \citep{riley2001steady}. 
The experimental study of steady streaming around a circular cylinder began with \cite{carriere1929analyse} and \cite{andrade1931circulations}, who examined periodically oscillating air around a cylinder and reported the now familiar quadrupole-like flow.
\cite{schlichting1932berechnung} carried out the first asymptotic analysis of steady streaming around a cylinder, and he compared the solution with his experiments on a vibrating cylinder in water. 
In the small amplitude limit, he matched the inner boundary layer with the outer potential flow, and his solution became the basis for future analysis. For example, \cite{wang1968high} adapted the layer analysis to include the curvature of the boundary, which is relevant at higher order. \cite{riley1965oscillating} and \cite{stuart1966double} analyzed the problem at high Reynolds number where a second boundary layer exists, and their results agreed well with Schlichting’s experiments. Finally, \cite{holtsmark1954boundary} analyzed and solved the flow everywhere as a regular perturbation problem. 

Though the problem of steady streaming has been studied extensively, past works have primarily focused on single-frequency oscillations, with comparatively less attention to multi-frequency oscillations \citep{davidson1972jets,kotas2008steady}. \cite{davidson1972jets} analyzed steady streaming of multi-frequency oscillation around a cylinder and found the resultant flow is the superposition of steady streaming flows induced by each single-frequency oscillation. 
Similarly, \cite{kotas2008steady} experimentally examined the steady flow around a sphere made to oscillate with two frequencies, and they found that the observed flows  were the sum of the streaming flows from the respective single-frequency oscillations.
The experiments in both of these works were carried out for small amplitude motion in which the streaming velocity is small, and the asymptotic analysis was only performed up to second order in amplitude. There are reasons to believe that the imposition of multiple oscillatory frequencies might break the spatial symmetry and induce net flow in one direction at higher order in amplitude. 

This work is inspired by recent experiments on a different physical system: an object sliding on a surface, undergoing two-frequency lateral oscillation \citep{hashemi2022net}. Specifically, the surface displacement was $(\ell/2)[\sin(\omega t) + \sin(\alpha\omega t)]$, where  $\ell$ is the amplitude, $\omega$ is the frequency, and $\alpha$ is a ratio of the two frequencies. For particular $\alpha$-values, the object exhibited a net translation.
To help understand the experimental observations, they analyzed a model in which the only force considered is the surface contact described by Coulomb's friction law. Representing the frequency ratio $\alpha$ as $p/q$, they derived a necessary condition for net motion: one of $p$ or $q$ is odd and the other is even, and validated this prediction with experiments.  For example, $\alpha = 2$, $3/2$, and $1/2$ showed net motion while $\alpha = 1$, $3$, and $5/3$ did not. We note the same necessary conditions for directed motion were derived earlier by \cite{reznik2001c} using a simpler friction model \citep{reznik1997analysis}.
Additional analysis and experiments for oscillations with different phase and amplitude were carried out in \cite{zhang2024theoretical}, and \cite{hui2024vibrational} further corroborated the theory with experiments with granular media. 

In this work, we use computations and analysis to examine steady streaming around a cylinder, whose oscillation is the sum of two sinusoids of different frequencies. 
We show that a cylinder vibrating with two-frequency oscillation leads to pumping, i.e.\ a streaming flow with net directed motion of the surrounding fluid.
We apply a small amplitude analysis to study steady streaming for two-frequency oscillation at low Reynolds numbers. 
Indeed, at second order in amplitude, steady streaming is a superposition of streaming due to individual frequencies, consistent with prior observations; however, pumping is a higher order effect and only occurs for certain frequency pairs.
Obtaining expressions for higher order terms is analytically intractable, but the form of the regular perturbation analysis allows us to deduce the structure of the solution and obtain necessary conditions for pumping.   

The sections are organized as follows. 
In Section 2, we introduce the motion of the cylinder and the nondimensionalization of the fluid equations.     
In Section 3, we present numerical simulations for the frequency ratio two and compare the flows with those of the single-frequency case. 
We measure the flux for varying amplitude, and we observe pumping is a third order effect.
In Section 4, we examine the asymptotic analysis of the problem at low amplitude. 
Even though we cannot solve the equations at third order, we show that a steady solution exists at third order and involves a net force responsible for pumping. 
Then, in Section 5, we expand the analysis to general frequency ratios and give necessary conditions for pumping, and we predict the order in amplitude at which pumping occurs.   
Finally, we confirm the theoretical results with computational simulations for select frequency ratios. 

\section{Problem Statement} \label{sec:Section1}
\subsection{Cylinder Motion}
We use two related formulations of the problem: one is in which the cylinder moves and the other in which the cylinder is stationary and the far\--field flow oscillates. For the moving cylinder formulation, a cylinder of radius $R$ and center $(X(t),0)$ moves in a 2D viscous fluid. The horizontal position of the cylinder's center is prescribed as
\begin{equation}
    X(t) = \frac{A}{2}[\sin(\Omega t) + \sin(\alpha\Omega t)],
    \label{Eq:PositionOfCylinder}
\end{equation}
where $A$ is the amplitude of oscillation, $\Omega$ is the angular frequency of the base oscillation, and $\alpha \geq 1$, the ratio of the two frequencies. The frequency ratio is assumed rational so that the flow is time periodic, and, if we choose $\alpha = 1$, we obtain the single-frequency case.

For analysis it is convenient to consider the cylinder fixed at the origin and impose a horizontal flow at infinity of strength  
\begin{equation}
    U(t) = -\frac{A\Omega}{2}[\cos(\Omega t) + \alpha\cos(\alpha \Omega t)].
    \label{eq:VelocityAtInfinity}
\end{equation}
Furthermore, no slip and no penetration will be imposed, and the boundary conditions in 2D, cylindrical coordinates are 
\begin{equation}
    u_r(r = R) = 0, \quad u_\theta(r = R) = 0.
    \label{eq:NoSlip}
\end{equation}

The two formulations are related by a change in reference frame.
Switching our frame of reference from the inertial frame where the cylinder is oscillating to the non\--inertial frame in which the cylinder is fixed results in a fictitious body force \citep{batchelor2000introduction}, 
\begin{equation}
    F = -\rho U'(t) \boldsymbol{e_x},
    \label{eq:BodyForce}
\end{equation}
where $\rho$ is the mass density and $\boldsymbol{e_x}$ is the direction vector in the horizontal direction. 
In the non\--inertial frame, we incorporate the fictitious body force \eqref{eq:BodyForce} in the  modified pressure defined as, 
\begin{equation*}
    \tilde{p} = p + \rho xU'(t).
\end{equation*}
For simplicity, we denote the modified pressure as $p$ below. 

\subsection{Nondimensionalization}
The equations are nondimensionalized using characteristic length scale $R$, time scale $\Omega^{-1}$, and velocity scale $R\Omega$. The flow around the cylinder is governed by the incompressible Navier Stokes equations:
\begin{gather}
	\label{Eq:NonDimMomentumEquation}
	\text{Re}\left(\boldsymbol{u_t} + \boldsymbol{u}\bcdot\bnabla\boldsymbol{u}\right) = -\bnabla p + \Delta \boldsymbol{u},\\
	\bnabla\bcdot\boldsymbol{u} = 0,
\end{gather}
where the Reynolds number is 
\begin{equation*}
\text{Re} = \frac{R^2\Omega}{\nu}, 
\end{equation*}
and $\nu$ is the kinematic viscosity. 
The boundary conditions \eqref{eq:VelocityAtInfinity} and \eqref{eq:NoSlip} become 
\begin{equation}
    u_r(r = 1) = 0, \qquad u_\theta(r = 1) = 0,
    \label{eq:NoSlipNonDim}
\end{equation}
and the horizontal flow at infinity is 
\begin{equation}
    U(t) = -\frac{\epsilon}{2}\left[\cos( t) + \alpha\cos(\alpha t)\right],
    \label{eq:NonDimVelocityAtInfinity}
\end{equation}    
where
\begin{equation*}
 \epsilon = \frac{A}{R}
\end{equation*}
is the dimensionless amplitude of oscillation. 

The Reynolds number, Re $= R^2\Omega/\nu$, is based on the length scale of the cylinder’s radius and corresponds to the boundary layer in which viscous forces remain relevant. The streaming Reynolds number, $\text{Re}_s$, is based on the length scale of the oscillation amplitude and it characterizes the Reynolds number associated with the streaming flow. The two Reynolds numbers are related by
\begin{equation*}
   \text{Re}_s = \frac{A^2\Omega}{\nu} = \epsilon^2 \text{Re}.
\end{equation*}
The flow structure depends on both $\text{Re}$ and $\text{Re}_s$. We refer to Figure 1 from \cite{chong2013inertial} (which is adapted from Figure 1 of \cite{wang1968high}) to illustrate the different flow regimes and associated analyses. Briefly, for high Reynolds numbers the flow exhibits a boundary layer \citep{schlichting1932berechnung,wang1968high}, and at high streaming Reynolds numbers, the streaming flow itself develops a second boundary layer \citep{stuart1966double,riley1965oscillating}. At low Reynolds number, there are no boundary layers, and the problem can be analyzed as a regular perturbation \citep{holtsmark1954boundary}. 
Though as noted in \cite{chong2013inertial}, the regular perturbation solution from \cite{holtsmark1954boundary} contains the single boundary layer solutions from \cite{schlichting1932berechnung} and \cite{wang1968high}, and is thus appropriate provided that the streaming Reynolds number is small.  
In this paper, we focus on the low streaming Reynolds number regime $(\text{Re}_s \ll 1)$ so that the problem can be analyzed as a regular perturbation, which allows us to compute the structure of solutions beyond second order needed to understand pumping from two-frequency oscillations.

\section{Computational Studies of Frequency Ratio 2} \label{sec:Section2}
We begin by examining the flow patterns of streaming for frequency ratio $\alpha = 2$ using numerical simulations. The numerical methods are described in Appendix \ref{sec:NumericalMethods}. We consider other frequency ratios in Section \ref{sec:Section5} where we show that $\alpha = 2$ yields the strongest pumping and the difference of the flow pattern from the single-frequency case is subtle for other frequency ratios.

\subsection{Comparing Flow Patterns} \label{sec:Section2.1} 
\begin{figure}
    \centering
    \includegraphics[width=1\linewidth]{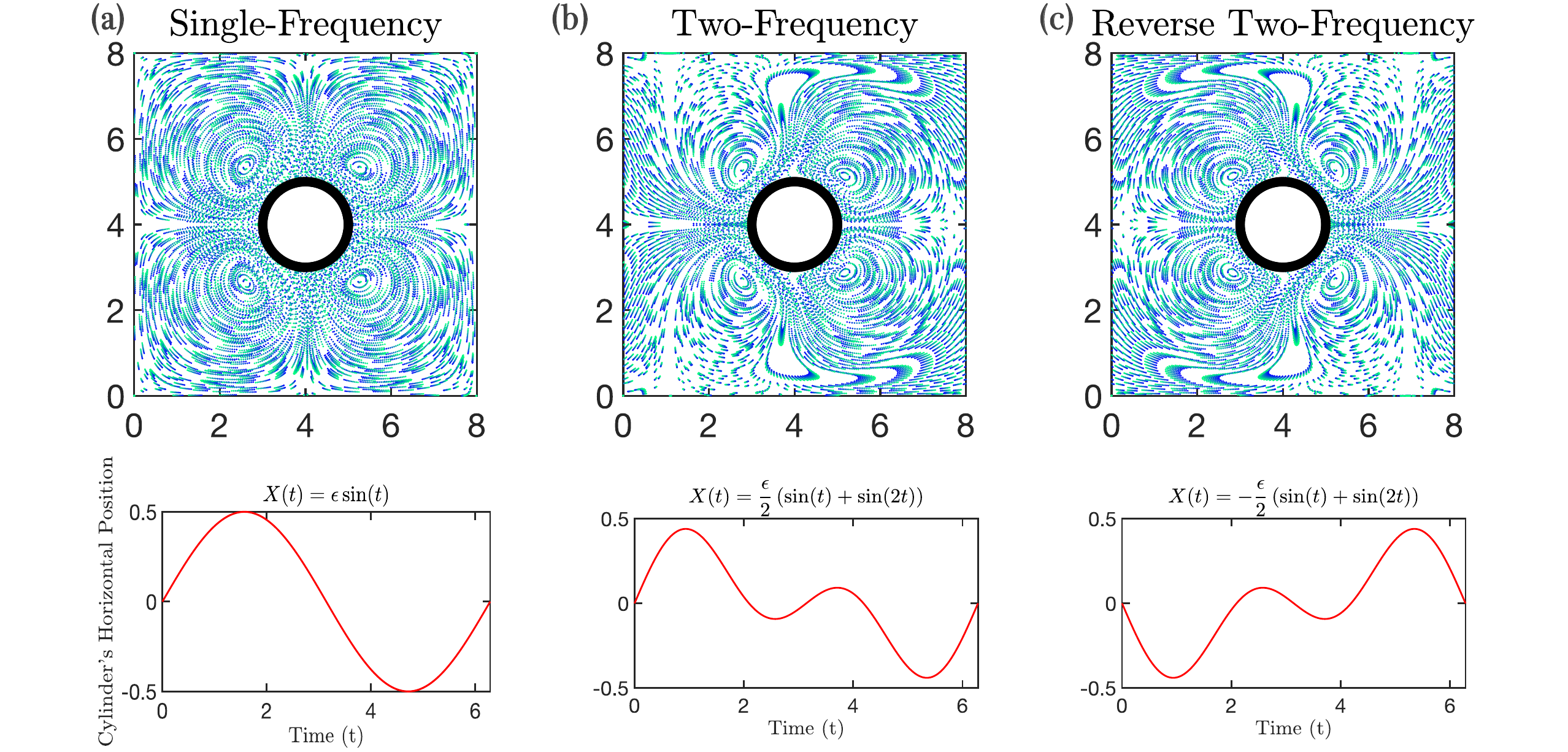}
    \caption{Streaming flows at time $T = 250$ for amplitude $0.5$ and Reynolds number $10$ for (a) single frequency motion, (b) two-frequency motion, and (c) two-frequency motion with time reversal. Shown are the positions of passive tracer particles over $10$ periods where the current location is colored green and the location $10$ periods prior is colored blue. The domain is a $8$ by $8$ with periodic boundary conditions. The bottom panels show the horizontal position of the cylinder's center over one period.}
    \label{fig:alpha=0_5}
\end{figure}

We place the cylinder in a square 8 by 8 domain with periodic boundary conditions and
solve for the flow at Re $= 10$. We solve for 250 periods which is sufficiently long to allow the initial transient to decay, and the resulting flows are visualized by the positions of passive marker particles from the previous 10 periods of time. 
As expected for the single-frequency case in Figure \ref{fig:alpha=0_5}(a), the flow resembles the classical, four vortices associated with steady streaming.  However, for two-frequency motion, shown in Figures \ref{fig:alpha=0_5}(b,c), the vortices are no longer symmetric and there is a net horizontal flow.
The bottom panels show the cylinder's horizontal position for each case. 
In Figure \ref{fig:alpha=0_5}(b), the net flow is left-to-right \--- away from the cylinder, the green particles (most current period) are to the right of the blue particles (past period), indicating a net rightward flow.
See movie 1 in the supplementary materials to compare the flow patterns of single-frequency, two-frequency, and reverse two-frequency oscillations. 
We note that because of the periodic boundary conditions these simulations correspond to a lattice of oscillating cylinders.

Figure \ref{fig:alpha=0_5}(c) shows the flow resulting from the time-reversed, two-frequency motion of the cylinder. In this case, the flow is moving right-to-left, and the flow pattern appears to be the reflection about the vertical axis of the flow pattern from Figure \ref{fig:alpha=0_5}(b).
The waveforms of the cylinders' positions for all three cases are depicted in the bottom row of Figure \ref{fig:alpha=0_5}.
Note that the time-reversed waveform of the two-frequency oscillation is not a phase shift of the original, as in the single-frequency case.  
For the time-reversed oscillation, the flow is moving right-to-left, and the flow pattern appears to be the reflection about the vertical axis of the flow pattern from Figure \ref{fig:alpha=0_5}(b). This result demonstrates that the direction of the net flow is determined by the time asymmetry of the cylinder's motion. In the discussion, we elaborate on how the time symmetry of the cylinder motion for different frequency ratios is related to pumping. We note that a similar result was observed in the experiments of \cite{hashemi2022net} and \cite{hui2024vibrational}.
When they reversed the polarity of the motion, the object translated in the opposite direction.

\subsection{Pumping in a Channel} \label{sec:Section2.2}
\begin{figure}
    \centering
    \includegraphics[width=1\linewidth]{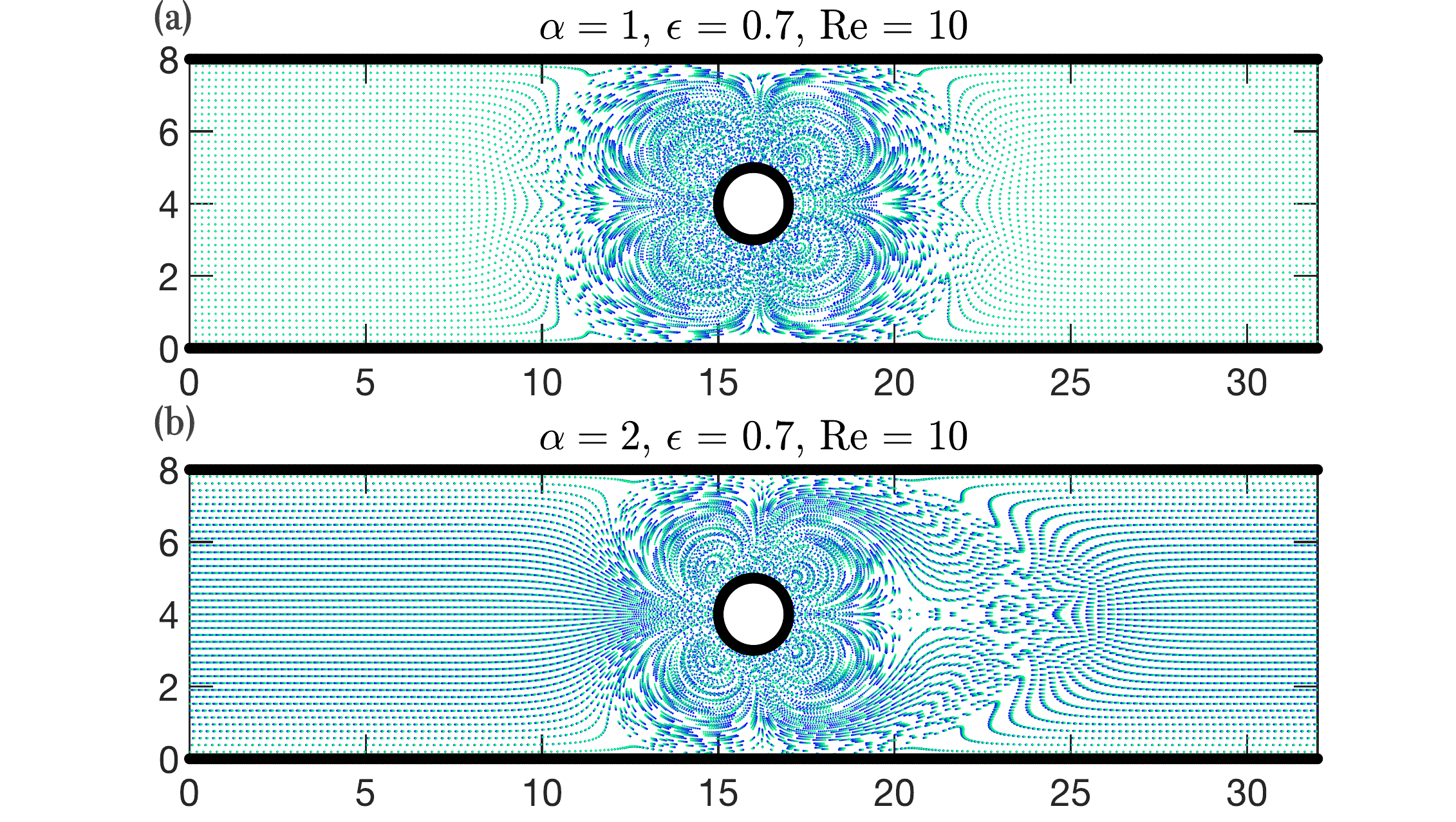}
    \caption{Streaming flows in a $32$ by $8$ channel for (a) single frequency ($\alpha = 1$) and (b) two-frequency ($\alpha = 2$) oscillations at time $T = 250$ visualized with amplitude $\epsilon = 0.7$ and Reynolds number $10$. Passive tracer particles highlight the positions of the flow from the last $10$ periods. Shown are the positions of passive tracer particles over $10$ periods where the current location is colored green and the location $10$ periods prior is colored blue.}
    \label{fig:channelAmp=0_3}
\end{figure}

The results in \ref{fig:alpha=0_5}(b) suggest that cylinders oscillating with two frequencies can be used to pump fluids. To test that idea more directly, we repeated the computational experiments in a channel of height 8 and length 32 with no-slip boundary conditions on the top and bottom and periodic conditions in the horizontal direction.
Figure \ref{fig:channelAmp=0_3} is a picture of movie 2 at time $T = 250$, and they show the contrast of single\-- and two\--frequency motions inside a channel.
Again, for the single-frequency case, the four vortices align symmetrically around the cylinder, and particles away from the cylinder remain stationary.  However, when the cylinder oscillates with two-frequencies, the vortices lose their symmetry, and fluid is pumped to the right. Applying the negative waveform again reversed the direction of motion (results not shown).   

\begin{figure}
    \centering
    \includegraphics[width=0.8\linewidth]{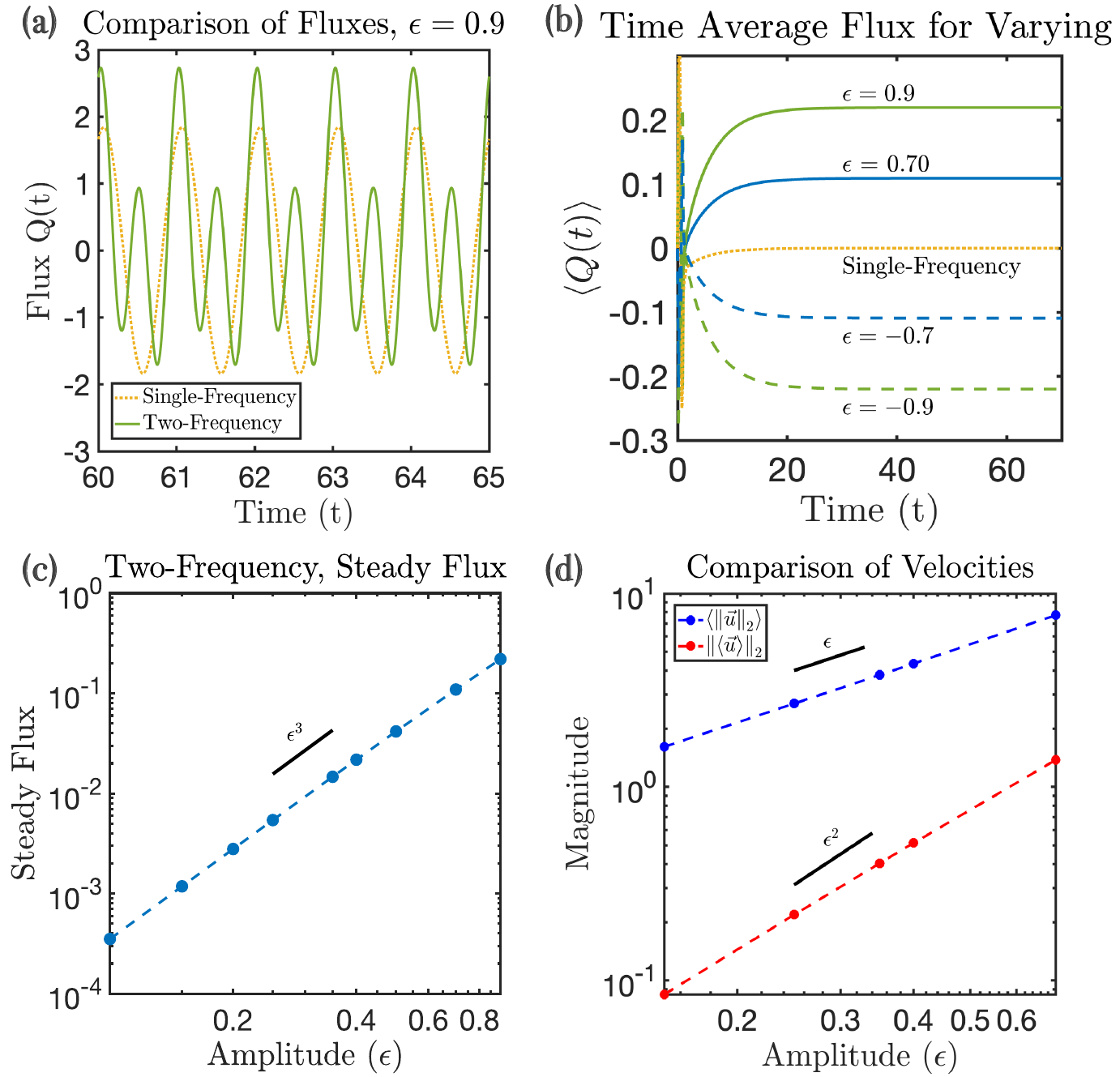}
    \caption{(a) Flux vs.\ time is presented for both the single\-- and two\--frequency case for amplitude 0.9 and Reynolds number 10. (b) Time-averaged fluxes are shown for varying amplitudes $(\epsilon$). (c) The steady time-averaged flux is third order in amplitude for two-frequency oscillations. (d) The size of the oscillatory component, $\langle\|u\|_2\rangle$, is first order in amplitude, and the size of the steady streaming flow, $\|\langle u\rangle\|_2$, is second order in amplitude for the two-frequency oscillation.}
    \label{Fig:ThirdOrderEffect}
\end{figure}

Figure \ref{Fig:ThirdOrderEffect}(a) shows the flux through a channel of height $H$, 
\begin{equation*}
    Q(t) = \int_{0}^{H} u(x, y, t)\;dy,
\end{equation*}
over $5$ periods for both the single\-- and two\--frequency cases for amplitude $\epsilon=0.9$. The flux for the two-frequency case shows a nonzero average flux on the scale of about 10\% of the amplitude of the oscillation. We quantify pumping by the time average of the flux over the period $T$:
\begin{equation}
    \langle Q(t)\rangle \;= \frac{1}{T}\int_{t - T}^{t} Q(\tau)\; d\tau.
    \label{eq:Flux}
\end{equation}
Figure \ref{Fig:ThirdOrderEffect}(b) shows the period-averaged flux as a function of time for different amplitudes. For each amplitude, the period-averaged flux approaches a steady value which depends on the amplitude. The time-reversed oscillation results in the negative of the flux for the non-reversed oscillation of the same amplitude.  

Figure \ref{Fig:ThirdOrderEffect}(c) shows that the steady flux grows like amplitude cubed. For single-frequency oscillations the magnitude of the steady flow is second order in amplitude and the oscillatory component is first order \citep{holtsmark1954boundary, wang1968high, chong2013inertial}. In Figure \ref{Fig:ThirdOrderEffect}(d) we show that for the two-frequency case these same scalings for the size of the steady and oscillatory flows result as measured by the $2$-norm of the period-averaged flow and the time average of the $2$-norm of the velocity, respectively. Previously, analysis on steady streaming has only been carried out to second order in amplitude, and so to derive the terms responsible for pumping, we need to extend the analysis to third order.

\subsection{Effect of Reynolds Number on Pumping}
\begin{figure}
    \centering
    \includegraphics[width=0.43\linewidth]{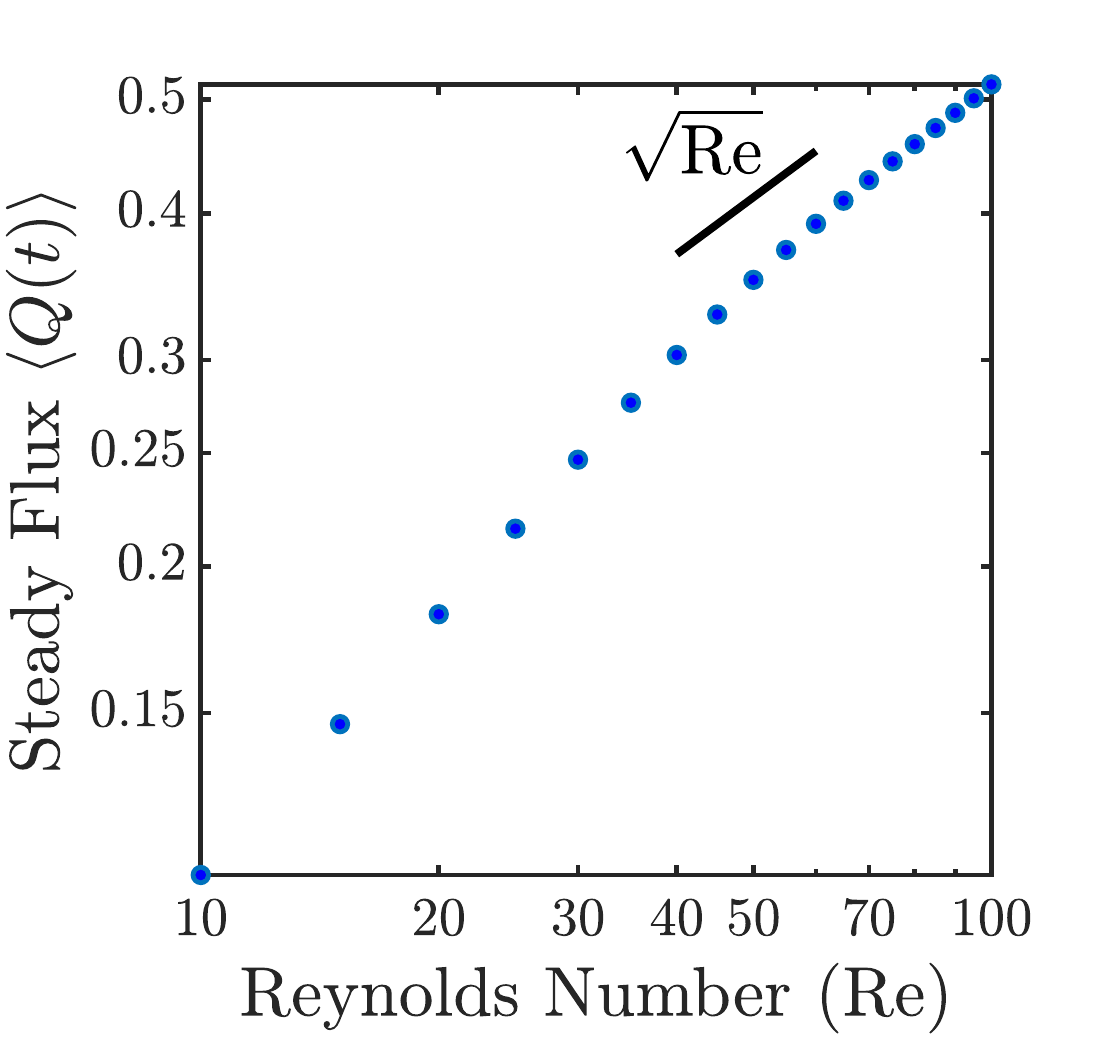}
    \caption{At amplitude $\epsilon = 0.7$, frequency ratio $\alpha = 2$, and time $\text{T} = 100$, the time-averaged fluxes for a 32 by 8 channel are measured for various Reynolds numbers. Beyond the critical Reynolds number (Re $\approx 40$) at which a boundary layer forms, the time\--averaged fluxes appear to scale as $\sqrt{\text{Re}}$.}
    \label{Fig:ReynoldsNumber1}
\end{figure}
We repeat the computational studies from the previous section to examine how the flux and flow structure change with Reynolds numbers.
We note that at $\text{Re} = 0$, the time\--averaged flow is zero because of the dynamic reversibility of Stokes flow, and thus there is no pumping at zero Reynolds number. 
Figure \ref{Fig:ReynoldsNumber1} shows the time\--averaged flux for $10\leq \text{Re}\leq 100$ at amplitude $\epsilon = 0.7$, frequency ratio $\alpha = 2$, and time $\text{T} = 100$.
The flux grows with Reynolds number and shows a change in scaling around $\text{Re} = 40$, above which the time\--averaged flux scales approximately as $\sqrt{\text{Re}}$.
This change in scaling coincides with the Reynolds number at which a boundary develops in the single\--frequency case \citep{chong2013inertial}.
\begin{figure}
    \centering
    \includegraphics[width=1\linewidth]{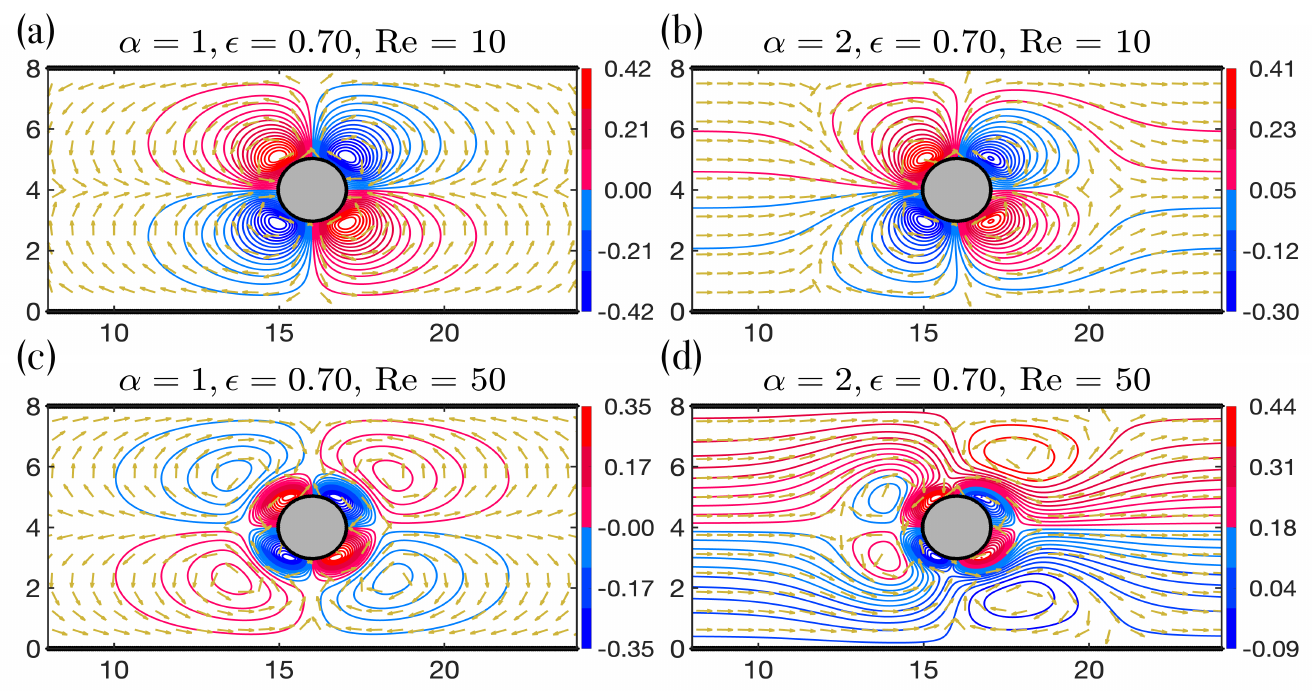}
    \caption{For a 32 by 8 channel at amplitude $\epsilon = 0.7$ and time $\text{T} = 100$, the streamlines and the flow directions (normalized velocity field) of the time\--averaged flow are shown at frequency ratios $\alpha = 1$ (a, c) and $\alpha = 2$ (b, d) and $\text{Re} = 10$ (a, b) and $\text{Re} = 50$ (c, d).}
    \label{Fig:ReynoldsNumber2}
\end{figure}

In Figure \ref{Fig:ReynoldsNumber2}, we show the streamlines and normalized velocity vectors of the time\--averaged flow for both single and two-frequency cases at Re = 10 and Re = 50 which are above and below the critical Reynolds number at which a boundary layer forms. At Re=10 for the single-frequency case we see the classical quadrupole-like structure where the red streamlines indicate counterclockwise flow, and the blue indicates clockwise flow (Figure \ref{Fig:ReynoldsNumber2}(a)).
For frequency ratio $\alpha = 2$, there is a clear flow to the right, and vortices are no longer left-right symmetric.  At Re = 50, there is a boundary layer present for both the single and two\--frequency cases consisting of four rotating vortices with counter-rotating vortices outside the layer (Figure \ref{Fig:ReynoldsNumber2}(c\--d)). For two\--frequency oscillations the net flow passes between the inner and outer vortices downstream from the cylinder.

\section{Low Amplitude Analysis} \label{sec:Section4}
We now analyze the problem in the limit of small amplitude $(\epsilon \ll 1)$, and we restrict our analysis to small streaming Reynolds numbers $(\text{Re}_s \ll 1)$ to apply a regular perturbation.
The single-frequency solution has been previously computed through second order in amplitude \citep{holtsmark1954boundary, chong2013inertial}, but extending the analysis to third order is not analytically tractable. However, we derive the structure of the higher order solutions, and, from the structure alone, we can understand why pumping is a third order effect for the two-frequency case. 

\subsection{Governing Equations}
We analyze the problem in terms of the stream function, $\psi$, which is related to the velocity by
\begin{equation}
    u_r = \frac{1}{r}\frac{\partial\psi}{\partial \theta}, \quad u_\theta = -\frac{\partial\psi}{\partial r},
    \label{eq:StreamFunctionRep} 
 \end{equation}
and
\begin{equation}
  \bnabla \times \boldsymbol{u} = -\Delta\psi.
\end{equation}
We take the curl of the momentum equation \eqref{Eq:NonDimMomentumEquation} to obtain
\begin{equation}
  \label{Eq:NonDimNavier1}
  \left(\Delta - Re\frac{\partial}{\partial t}\right)\Delta \psi =  Re\;\boldsymbol{u}\bcdot\bnabla\left(\Delta\psi\right).
\end{equation}
We solve the problem in the reference frame in which the cylinder is stationary with an oscillatory flow at infinity. The boundary conditions in terms of the stream function are
\begin{gather}
    \label{Eq:NonDimNavier2}
    \psi(r = 1) = 0, \quad \frac{\partial\psi}{\partial r}(r = 1) = 0, \\
    \label{Eq:NonDimNavier3}
    \psi \sim -\frac{\epsilon r}{2}\left(\cos(t) + 2\cos(2t)\right)\sin(\theta), \quad r \rightarrow \infty.
\end{gather}
We consider the low amplitude limit and expand the solution in terms of $\epsilon = A/R$:
$$\psi = \epsilon\psi_1 + \epsilon^2\psi_2 + \epsilon^3\psi_3 + O\left(\epsilon^4\right).$$
Substituting the expansion into \eqref{Eq:NonDimNavier1}, \eqref{Eq:NonDimNavier2}, and \eqref{Eq:NonDimNavier3}, we obtain the successive equations at each order:
\begin{gather}
  \textbf{First Order} \nonumber  \\
  \left(\Delta - \text{Re}\frac{\partial}{\partial t}\right)\Delta \psi_1 = 0,   \label{Eq:FirstOrderProblem1} \\   
  \psi_1(r = 1) = 0, \qquad \frac{\partial\psi_1}{\partial r}(r = 1) = 0, \label{Eq:FirstOrderProblem2} \\
  \psi_1 \sim -\frac{r}{2}\left(\cos(t) + 2\cos(2t)\right)\sin(\theta), 
     \quad r \rightarrow \infty \label{Eq:FirstOrderProblem3}
\end{gather}

\begin{gather}
  \nonumber
  \textbf{Second Order} \nonumber \\
  \left(\Delta - \text{Re}\frac{\partial}{\partial t}\right)\Delta \psi_2 = \text{Re}\; 
      \boldsymbol{u_1}\bcdot\bnabla\left(\Delta\psi_1\right), \label{Eq:SecondOrderProblem1} \\
  \psi_2(r = 1) = 0, \quad \frac{\partial\psi_2}{\partial r}(r = 1) = 0, \label{Eq:SecondOrderProblem2} \\
  \psi_2 \sim 0, \quad r \rightarrow \infty, \label{Eq:SecondOrderProblem3}
\end{gather}

\begin{gather}
  \textbf{Third Order}\nonumber\\
  \left(\Delta - \text{Re}\frac{\partial}{\partial t}\right)\Delta \psi_3 = \text{Re}\; 
     \boldsymbol{u_1}\bcdot\bnabla\left(\Delta\psi_2\right) + \text{Re}\; \boldsymbol{u_2}\bcdot\bnabla\left(\Delta\psi_1\right),
     \label{Eq:ThirdOrderProblem1} \\
  \psi_3(r = 1) = 0, \quad \frac{\partial\psi_3}{\partial r}(r = 1) = 0, \label{Eq:ThirdOrderProblem2} \\
  \psi_3 \sim 0, \quad r \rightarrow \infty.   \label{Eq:ThirdOrderProblem3}
\end{gather}

\subsection{Solution Structure}  

\subsubsection{Single-Frequency Oscillation}
\label{Sec:Solution_Structure:Single-Frequency_Oscillation}
Before we consider the two-frequency case, we first examine the solution structure for single-frequency oscillation. The only difference in the equations for the single-frequency and two-frequency cases is the boundary condition at infinity at first order. For the single-frequency case, equation \eqref{Eq:FirstOrderProblem3} is replaced by 
\begin{equation}
\psi_1 \sim -r\cos(t)\sin(\theta).\nonumber
\end{equation}
This boundary condition determines the form of the solution at first order as
\begin{equation}
 \psi_1 = \Re\left(a_1(r)e^{-it}\right)\sin(\theta),
 \label{Eq:SingleFirstOrderSolution}
\end{equation}
where $\Re\left(z\right)$ denotes the real part of $z$.
Analytically, $a_1(r)$ is a solution composed of Hankel functions \citep{watson1922treatise}.
We substitute \eqref{Eq:SingleFirstOrderSolution} into the quadratic nonlinearity on the right side of Equation \eqref{Eq:SecondOrderProblem1} to determine the form of the solution at second order. This computation involves all products of $e^{-it}$ and $e^{it}$, which results in a steady term and terms proportional to $e^{\pm i2t}$. Therefore, at second order, we obtain 
\begin{equation}
  \psi_2 = \Re\left(b_2(r)e^{-2it}\right)\sin(2\theta) + b_0(r)\sin(2\theta).
  \label{Eq:SingleSecondOrderSolution}
\end{equation}
Equation \eqref{Eq:ThirdOrderProblem1} at third order involves products of \eqref{Eq:SingleFirstOrderSolution} and  \eqref{Eq:SingleSecondOrderSolution} on its right-hand side. This computation involves products of $e^{\pm it}$ with $\{1,e^{\pm 2it}\}$, which results in terms proportional to $e^{\pm i3t}$ and  $e^{\pm it}$. The third order solution is of the form 
\begin{equation}
 \psi_3 = \Re\left(c_3(r)e^{-i3t} + c_1(r)e^{-it}\right)\sin(3\theta) + \Re\left(d_3(r)e^{-i3t} + d_1(r)e^{-it}\right)\sin(\theta).
 \label{Eq:SingleThirdOrderSolution}
\end{equation}
Therefore, for the single-frequency case, there is no steady term at third order. 
\subsubsection{Two-Frequency Oscillation}
For the two-frequency case, we obtain 
\begin{equation}
    \psi_1 = \Re\left(a_1(r)e^{-it} + a_2(r)e^{-2it}\right)\sin(\theta),
    \label{Eq:TwoFirstOrderSolution}
\end{equation}
as the solution to \eqref{Eq:FirstOrderProblem1} \-- \eqref{Eq:FirstOrderProblem3}.
As before, we substitute the solution at first order into the right side of equation \eqref{Eq:SecondOrderProblem1} to deduce the structure of the second order solution as
\begin{eqnarray}
\psi_2 & = & \Re\left(\sum_{k=1}^{4}b_k(r)e^{-ikt}\right)\sin(2\theta) + b_0^{(1)}(r)\sin(2\theta) + b_0^{(2)}(r)\sin(2\theta) \nonumber\\
&=& \Re\left(\left(b_{2}(r)e^{-2it}+ b_0^{(1)}(r)\right)
                    + \left(b_{4}(r)e^{-4it}+ b_0^{(2)}(r)\right) \right.\nonumber\\
&& \mbox{} + \left.\left(b_{1}(r)e^{-it}+ b_{3}(r)e^{-3it}\right)\right)\sin(2\theta).
\label{Eq:TwoSecondOrderSolution2}
\end{eqnarray}

We express the solution in this second form with three pairs of terms grouped together to highlight the origin of the different terms in the sum. The first two pairs involving a steady term and unsteady term correspond to the single-frequency solutions from each of the two driving frequencies, and the last pair of unsteady terms arises from the interaction between the driving frequencies. Thus as has been reported previously, the steady flow at second order is the superposition of two streaming fields caused by the individual, single-frequency oscillations \citep{davidson1972jets,kotas2008steady}. 

The nonlinear interaction between the two frequencies only affects the unsteady terms at second order, but these interactions produce a steady flow at third order. 
The solution at third order has the form 
\begin{eqnarray}
\psi_3 & = & \Re\left(\sum_{k=1}^6 c_k(r)e^{-ikt}\right)\sin(3\theta) + c_0(r)\sin(3\theta) \nonumber\\
&& \mbox{} + \Re\left(\sum_{k=1}^6 d_k(r)e^{-ikt}\right)\sin(\theta) + d_0(r)\sin(\theta).
\label{Eq:TwoThirdOrderSolution}
\end{eqnarray}
Unlike the single-frequency solution \eqref{Eq:SingleThirdOrderSolution}, there are steady terms at third order in amplitude. 
%
In the following sections, we explain how the third order, steady terms produce the observed left-right asymmetry and are responsible for producing a net force that pumps the fluid.

\subsection{Symmetry of Steady Flow}
To illustrate how the steady flows at third order create asymmetry, consider the steady solution,
\begin{equation}
    \psi^s = \epsilon^2 b(r)\sin(2\theta) + \epsilon^3 c(r)\sin(3\theta) + \epsilon^3 d(r)\sin(\theta) + O\left(\epsilon^4\right).
    \label{Eq:SteadySolution}
\end{equation}
We show that the familiar second order flow has a different spatial symmetry than the third order flow that arises from the interactions of the flows generated by the two different driving frequencies. 
Specifically, the spatial symmetry arises from the $\theta$-dependence of the stream functions.
To illustrate these symmetries, we examine the streamlines of the three stream functions,
\begin{equation}
\psi_2 = f(r)\sin(2\theta),\quad \psi_3^{(1)} = f(r)\sin(\theta),\quad \psi_3^{(3)} = f(r)\sin(3\theta),\nonumber
\end{equation}
where 
\begin{equation}
    f(r) = \frac{(r-1)^2}{1 + r^4}.
    \label{Eq:functionDefinition}
\end{equation}
This choice of $f(r)$ satisfies the boundary conditions on the cylinder and decays to zero at infinity, but it is not related to the solution of the equations.

\begin{figure}
    \centering
    \includegraphics[width=0.8\linewidth]{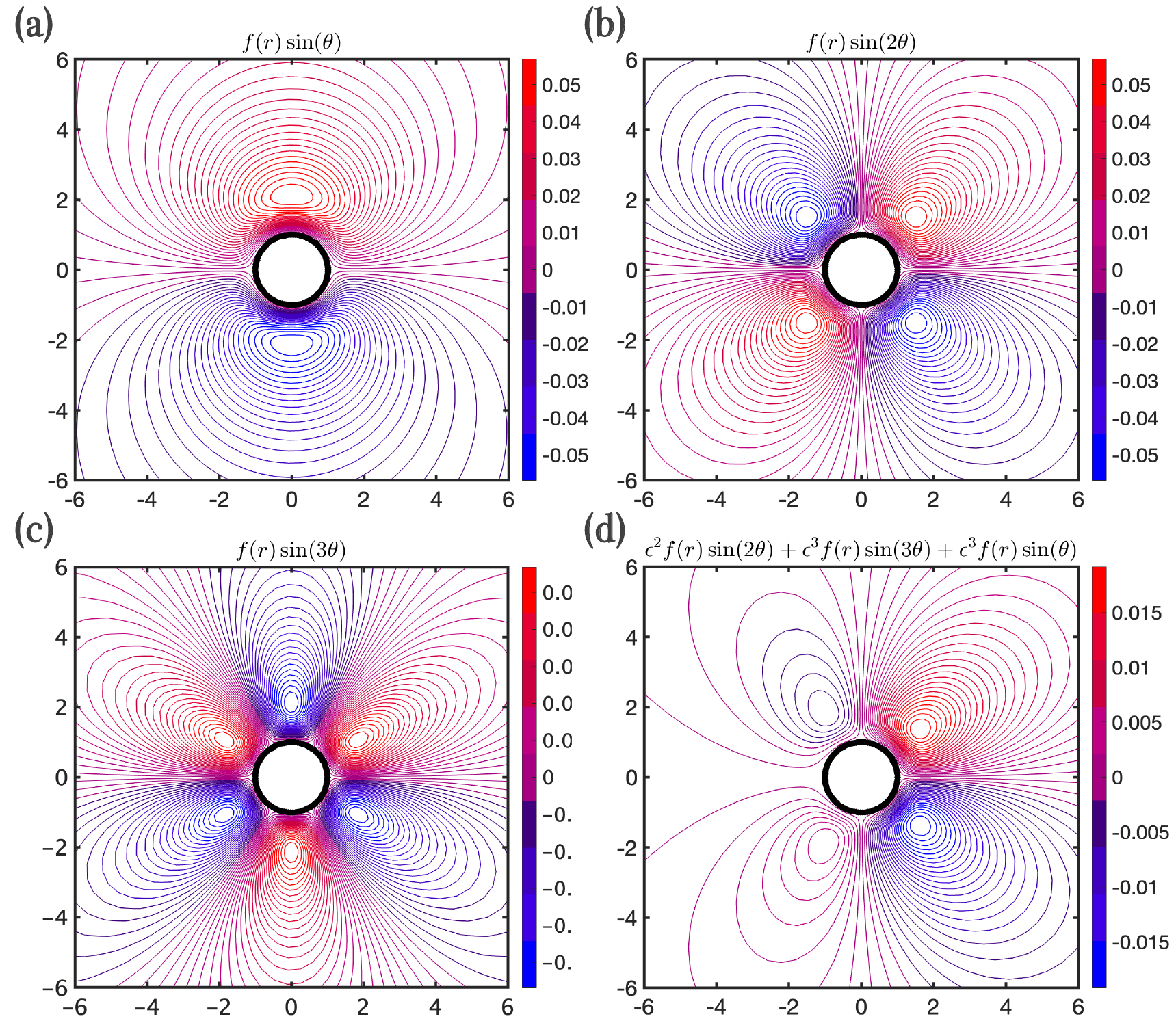}
    \caption{The contours of the stream functions are shown for (a) $\psi_2 = f(r)\sin(2\theta)$, (b) $\psi_3^{(1)} = f(r)\sin(\theta)$, and (c) $\psi_3^{(3)} = f(r)\sin(3\theta)$, where $f(r)$ is defined by \eqref{Eq:functionDefinition}. (d) Streamlines of the sum, $\psi = \epsilon^2 f(r)\sin(2\theta) + \epsilon^3 f(r)\sin(3\theta) + \epsilon^3 f(r)\sin(\theta)$, for $\epsilon = 0.45$ exhibit a left-right asymmetry.}
    \label{fig:SymmetryBreaking}
\end{figure}

The streamlines of $\psi_2, \psi_3^{(1)},$ and $\psi_3^{(3)}$ are shown in Figures \ref{fig:SymmetryBreaking}(a)-(c), while the streamlines of the weighted sum is presented in Figure \ref{fig:SymmetryBreaking}(d).
All the terms share a common up-down symmetry where the values at the points reflected across the horizontal axis are opposite in sign.   
Specifically, the stream functions are odd in $y$:
\begin{equation}
\psi_2(x, -y) =-\psi_2(x,y), \quad \psi_3^{(1)}(x, -y) = -\psi_3^{(1)}(x, y), \quad \psi_3^{(3)}(x, -y) = -\psi_3^{(3)}(x, y),\nonumber
\end{equation}
and the sum of the stream functions maintain this up-down symmetry.   
However, $\psi_2$ has a different left-right symmetry than the symmetry shared by the $\psi_3$'s.   
Specifically, $\psi_2$ is odd in $x$ while the $\psi_3$'s are even in $x$:
\begin{equation}
\psi_2(-x, y) =-\psi_2(x,y), \quad \psi_3^{(1)}(-x, y) = \psi_3^{(1)}(x, y), \quad \psi_3^{(3)}(-x, y) = \psi_3^{(3)}(x, y).\nonumber
\end{equation}
Therefore, the sum of the stream functions, illustrated in Figure \ref{fig:SymmetryBreaking}(d), retains the up-down symmetry, but lacks the left-right symmetry. 
%
%
\subsection{Net Force} \label{sec:Section4.4}
We compute the net force on the cylinder and show that the steady term proportional to $\sin(\theta)$ that appears at third-order for two-frequency oscillations indicates that there is a net force on the cylinder. Assume a steady solution of the form,
\begin{equation}
   \psi^s = \sum_{k = 1}^\infty f_k(r)\sin(k\theta). 
   \label{eq:TheoreticalStreamFunction}
\end{equation}
The pressure can be computed using the momentum equation \eqref{Eq:NonDimMomentumEquation} to obtain 
\begin{equation}
   p^s = p_\infty + \sum_{k=1}^\infty g_k(r)\cos(k\theta).
   \label{eq:TheoreticalPressure}
\end{equation}
The traction force in the horizontal direction is
\begin{equation}
    \left(\sigma\cdot \hat{e}_r\right)\cdot \hat{e}_x = \sigma_{rr}\cos(\theta) - \sigma_{r\theta}\sin(\theta),   
    \label{SurfaceStress}
\end{equation}
where $\hat{e}_{r}$ and $\hat{e}_x$ are the direction vectors for the $r$- and $x$-directions, respectively, and $\sigma$ is the stress tensor. Using the boundary conditions on the cylinder from equation \eqref{eq:NoSlipNonDim}, the components of the stress on the cylinder are
\begin{equation*}
    \sigma_{rr} = -p_{\infty} - \sum_{k=1}^{\infty}g_k(1)\cos(k\theta), \qquad \sigma_{r\theta} = -\sum_{k=1}^{\infty}f_k''(1)\sin(k\theta).
\end{equation*}
Integrating the surface stress \eqref{SurfaceStress} over the cylinder, all the terms corresponding to $k \geq 2$ vanish due to symmetry, and the net horizontal force is 
\begin{equation}
    \text{Net Force} = \int_0^{2\pi} \sigma_{rr}\cos(\theta) - \sigma_{r\theta}\sin(\theta)\;d\theta = \pi\left(-g_1(1) + f_1''(1)\right).
    \label{NetForceComp}
\end{equation}
Therefore the term proportional to $\sin(\theta)$ that appears at third order in equation \eqref{Eq:SteadySolution} contributes a net force which is responsible for the observed pumping.  

This force calculation gives insight as to why pumping does not occur for the single-frequency case. From Section  \ref{Sec:Solution_Structure:Single-Frequency_Oscillation}, the only steady term through third order in amplitude is proportional to $\sin(2\theta)$, and the only nonzero term in the pressure expansion, equation \eqref{eq:TheoreticalPressure}, is proportional to $\cos(2\theta)$. Therefore there is no net force for single-frequency oscillations through third order in amplitude, though we have not yet considered the possibility of a net force resulting at higher order for the single-frequency case. In the next section, we analyze the form of the solution at higher orders, and we extend the analysis to general frequency ratio to derive necessary conditions on the frequency ratio for pumping.

\section{Analysis of General Frequency Ratios}\label{sec:Section5} 
In the previous section, we showed that the steady stream function for frequency ratio $\alpha=2$ contained a term at third order proportional to $\sin(\theta)$. 
In this section, we examine the solution structure of $\psi_n$, the solution to the $n^{th}$ order perturbative equation \eqref{Eq:NthOrderEquation}, to determine whether there is potentially a net force on the fluid due to the presence of a $\sin(\theta)$ term in the expansion and if so, what order. 

For this analysis, it is convenient to express the motion of the cylinder as 
\begin{equation}
    X(t) = \frac{\epsilon}{2}\left(\sin(a t) + \sin(b t)\right),
    \label{Eq:TwoFrequencyMotion}
\end{equation} 
where gcd$(a, b) = 1$. This expression is a slight modification from equation \eqref{Eq:PositionOfCylinder}, and it amounts to a different definition of the Reynolds number than the one used in the previous sections. We again change reference frames by fixing the cylinder, and the governing equations have the form, 
\begin{gather}
    \label{Eq:TwoFrequency Problem}
    \left(\Delta - \text{Re}\frac{\partial}{\partial t}\right)\Delta \psi = \text{Re}\;\boldsymbol{u}\bcdot\bnabla\left(\Delta\psi\right), \\
    \label{Eq:TwoFrequency ProblemConditionCylinder}
    \psi(r = 1) = 0, \qquad \frac{\partial\psi}{\partial r}(r = 1) = 0, \\
    \label{Eq:TwoFrequency ProblemConditionInfinity}
    \psi \sim -\frac{\epsilon r }{2} \left(a\cos(a t) + b\cos(bt)\right)\sin(\theta), \quad r \rightarrow \infty.
\end{gather}
Expanding
\begin{equation}
\psi = \epsilon\psi_1 + \epsilon^2\psi_2 + \epsilon^3\psi_3 + O\left(\epsilon^4\right)\nonumber
\end{equation}
in the low amplitude limit and substituting this expansion into Equation \eqref{Eq:TwoFrequency Problem}, we obtain successive equations of the form
\begin{equation}
    \Delta\left(\Delta - \text{Re}\;\frac{\partial}{\partial t}\right) \psi_n = \sum_{i+j = n} \text{Re}\;\boldsymbol{u_i} \bcdot\bnabla(\Delta \psi_j).
    \label{Eq:NthOrderEquation}
\end{equation}
The boundary conditions on the cylinder at each order are
\begin{equation}
    \psi_n(r = 1) = 0, \qquad \frac{\partial\psi_n}{\partial r}(r = 1) = 0,
    \label{Eq:NthOrderEquationBdCylinder}
\end{equation}
and the asymptotic condition at $r=\infty$ is 
\begin{equation}
    \label{Eq:NthOrderEquationBdInfinity}
    \psi_n \sim
        \begin{cases}
            \frac{-r}{2}\left(a\cos(at) + b\cos(bt)\right)\sin(\theta) & n = 1, \quad r\rightarrow \infty\\
            \\ 
            0 & n > 1, \quad r\rightarrow \infty.
        \end{cases}
\end{equation}

The solution to \eqref{Eq:NthOrderEquation}\--\eqref{Eq:NthOrderEquationBdInfinity} is of the form
\begin{equation}
  \psi_{n} = \sum_{m=1}^{\infty}\sum_{k,m=-\infty}^{\infty} f^{n}_{k,m}(r)\sin\left(m\theta\right)e^{ikt},\nonumber
\end{equation}
but the number of nonzero terms is finite at each order. The boundary condition determines the nonzero terms at leading order, and at higher orders the structure of the solution depends on the products of the lower order terms. In the following sections, we first determine which terms proportional to $\sin\left(m\theta\right)$ are nonzero at each order, and then we derive conditions for the presence of a steady term ($k=0$). 

%
%
\subsection{$\theta$-dependent Terms}
From the form of the far-field boundary condition \eqref{Eq:NthOrderEquationBdInfinity}, the leading order solution is of the form
\begin{equation}
    \psi_1 = \Re\left(a_2(r)e^{-ibt} + a_1(r)e^{-iat}\right)\sin(\theta) = C_1(r,t)\sin(\theta).
    \label{Eq:TwoFrequencyFirstOrderSolution}
\end{equation} 
The form of the right-hand side of \eqref{Eq:NthOrderEquation} dictates the solution structure at higher orders. Substituting $\psi_1$ into the quadratic nonlinearity, the right-hand side of \eqref{Eq:NthOrderEquation} is proportional to  $\sin(\theta)\cos(\theta)$, and thus the solution is proportional to $\sin(2\theta)$. Therefore, $\psi_2$ is of the form
\begin{equation}
\begin{array}{ll}
\psi_2 &= \displaystyle\Re\left(b_4(r)e^{-i2bt} + b_3(r)e^{-i(a+b)t} + b_2(r)e^{-i2at} + b_1(r)e^{-i(a-b)t} + b_0(r)\right)\sin(2\theta)\\[10pt]
\displaystyle &= C_2(r,t)\sin(2\theta).
 \end{array}
 \label{Eq:TwoFrequencySecondOrderSolution}
\end{equation}
The right-hand side of \eqref{Eq:NthOrderEquation} for $\psi_3$ involves the products of derivatives of $\psi_1$ and $\psi_2$. We obtain terms involving $\sin(2\theta)\cos(\theta)$ and $\sin(\theta)\cos(2\theta)$. Using the trigonometric identities,
\begin{equation}
2\sin(2\theta)\cos(\theta) = \sin(3\theta) + \sin(\theta), \qquad 2\sin(\theta)\cos(2\theta) = \sin(3\theta) - \sin(\theta),\nonumber
\end{equation}
$\psi_3$ can be expressed as
\begin{equation}
\psi_3 = C_3(r,t)\sin(3\theta) + C_4(r,t)\sin(\theta).\nonumber
\end{equation}
Similarly, the solution of $\psi_4$ has the form
\begin{equation}
\psi_4 = C_5(r,t)\sin(4\theta) + C_6(r,t)\sin(2\theta).\nonumber
\end{equation}

The first four orders show that the solution for odd values of $n$ involve terms proportional to $\sin(m\theta)$ for odd value of $m$ through $n=m$, and similarly for even values of $n$. In Appendix \ref{sec:SectionA} we prove a lemma which shows that this pattern holds for all $n$ for a general multi-modal oscillation. Additionally we prove an analogous result for the pressure. From Section \ref{sec:Section4.4}, a net horizontal force requires a term in the steady stream function proportional to $\sin(\theta)$ or a term in the steady pressure proportional to $\cos(\theta)$. Putting these two results together leads to the following necessary condition for pumping:
\begin{result}    \label{thm:FirstTheorem}
  A necessary condition for pumping is the existence of an odd-valued $n$ such that $\psi_n$ has the steady component.
\end{result}
\noindent
In the next section we examine the frequencies that occur at each order, and derive conditions for pumping based on the frequency ratio. 

%
%
\subsection{Necessary Conditions for Pumping}\label{sec:NoPumpingSingle}

\subsubsection{No Pumping for Single Frequency}
We demonstrate that single-frequency oscillations do not pump by showing that $\psi_n$ cannot have a steady component when $n$ is odd. We examine the frequencies that arise at each order and deduce the orders at which the steady terms occur. For example, Figure 5 of \cite{willis2024quasi} illustrates how the frequencies at each order arise from the lower order frequencies for the single-frequency case. 

The structure of the leading order solution for the single-frequency case $(a = b = 1)$ is given in Equation \eqref{Eq:SingleFirstOrderSolution}.
Alternately, we can express the solution as
\begin{equation}
\psi_1 = G_1^{1}(r,\theta)e^{-it} + \overline{G_1^{1}(r,\theta)}e^{it},
\end{equation}
where 
\begin{equation}
G_1^{1}(r,\theta) = \frac{a_1(r)\sin(\theta)}{2}.
\end{equation}
Similarly, the second order solution \eqref{Eq:SingleSecondOrderSolution} is rewritten as 
\begin{equation}
\psi_2 = G_2^2(r, \theta)e^{-i2t} + \overline{G_2^2(r, \theta)}e^{i2t} + G_0^2(r, \theta),\nonumber
\end{equation}
where
\begin{equation}
G_2^2(r, \theta) = \frac{b_2(r)\sin(2\theta)}{2}, \qquad G_0^2(r, \theta) = b_0(r)\sin(2\theta).
\end{equation}
The time dependence of the individual terms in $\psi_2$, i.e., $e^{-2it}$, $e^{2it}$, and $1=e^{i0t}$, result from the products of $e^{-it}$ and $e^{it}$ through the quadratic nonlinearity of the right-hand side of (\ref{Eq:NthOrderEquation}).
Extending the pattern to higher order $n$, if $\psi_n$ has the term proportional to $e^{i\omega t}$, then 
$$e^{i\omega t} = \left(e^{-it}\right)^{n_1}\left(e^{it}\right)^{n_2},$$  
where $n_1 + n_2 = n$ and $0\leq n_1, n_2 \leq n$.

Now, we show that the steady component cannot appear when $n$ is odd. Using the fact $n_2 = (n - n_1)$, 
$$\omega = -n_1 + n_2 = -2n_1 + n.$$
For the steady term to appear $(\omega = 0)$, $n$ must be even. Combining this result with Result \ref{thm:FirstTheorem} shows that pumping does not occur for single-frequency oscillations.

\subsubsection{Two Frequencies}
The two-frequency case involves products of $e^{\pm iat}$ and $e^{\pm ibt}$ leading to a more complicated analysis. However, using a similar parity argument, we obtain the following results.   
\begin{result}
  \label{AntiperiodicTheorem}
   Suppose that $a$ and $b$ are both odd and $\psi_n$ is the solution of 
   \eqref{Eq:NthOrderEquation}\--\eqref{Eq:NthOrderEquationBdInfinity}. Steady terms only appear
   for $n$ even, and thus pumping will not occur. 
\end{result}
\begin{result}
  \label{NonAntiperiodicTheorem}
   Suppose that only one of $a$ or $b$ is even (and the other is odd) and $\psi_n$ is the solution of 
   \eqref{Eq:NthOrderEquation}\--\eqref{Eq:NthOrderEquationBdInfinity}. $\psi_n$ can have the steady 
   component for $n$ odd if $n \geq (a+b)$. Therefore the minimum order at which pumping occurs is order $a+b$.
\end{result}\noindent
See Appendix \ref{sec:Pumping_proofs} for the proofs of Results \ref{AntiperiodicTheorem} and \ref{NonAntiperiodicTheorem}. In Result \ref{NonAntiperiodicTheorem}, we give the smallest order of amplitude at which pumping can occur; however, we did not guarantee pumping. For simplicity, we defined the radial components as coefficients in the analysis, but these coefficients could equate to zero.

%
%
\subsection{Numerical Results for Other Frequency Ratios}\label{sec:OtherAlphaSim}
\begin{figure}
    \centering
    \includegraphics[width=1\linewidth]{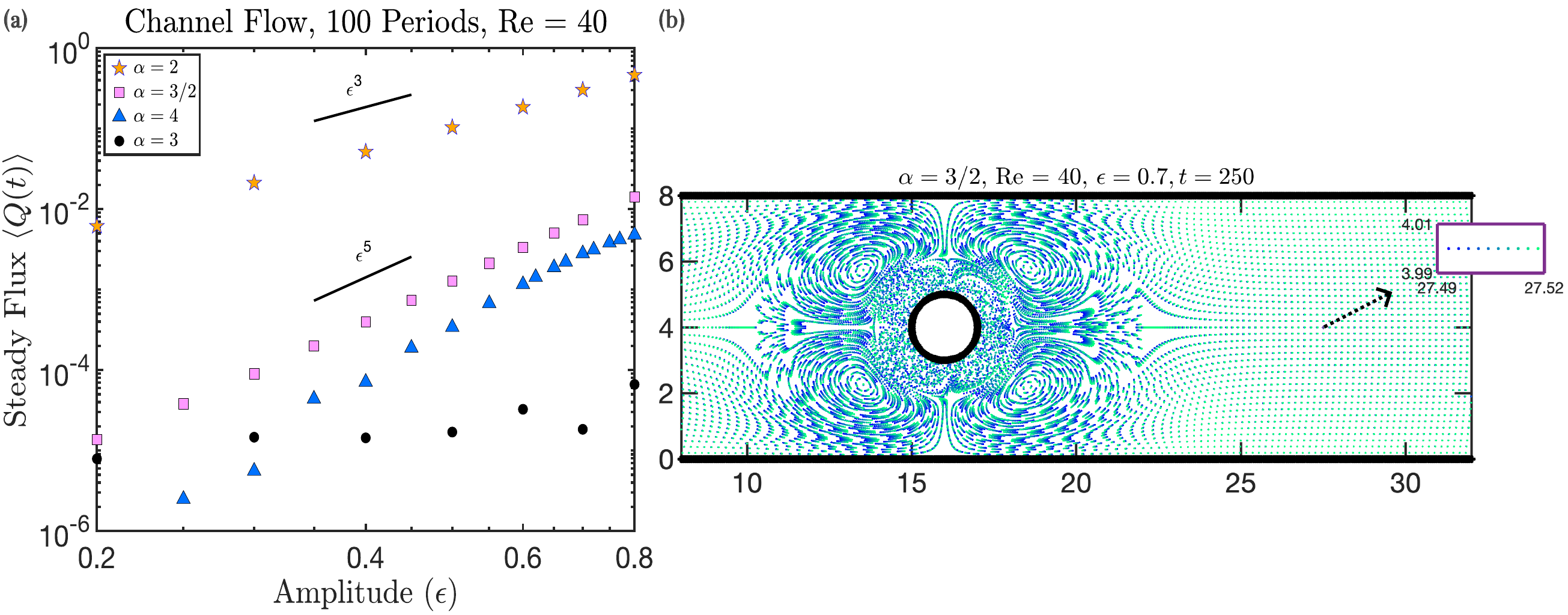}
    \caption{(a) Time-averaged flux vs.\ amplitude for frequency ratios $\alpha = 2$, $3/2$, $4$, and $3$. (b) Steady streaming in a channel for frequency ratio $\alpha = 3/2$ visualized at amplitude $0.7$ and Reynolds number $40$. Shown are the positions of passive tracer particles over $10$ periods where the current location is colored green and the location $10$ periods prior is colored blue. Though the fluid is being pumped, net motion is not obvious on this scale. The inset shows the region around a single tracer particle which confirms that the fluid is moving slowly downstream.}
    \label{Fig:FiftheOrderEffect}
\end{figure}

According to Result \ref{NonAntiperiodicTheorem}, for small amplitude $\epsilon$, the strength of pumping scales like $\epsilon^n$, where $n = a + b$ is the sum of the two frequencies and must be odd.
Thus, the strongest pumping results when $n = 3$, corresponding to frequency ratio $\alpha = 2$.
The next smallest order at which pumping can occur is fifth order. From Result \ref{NonAntiperiodicTheorem}, a fifth order scaling can only occur when the frequency ratio is $3/2$ or $4$, respectively, corresponding to $a = 2, b = 3$ or $a = 1, b = 4$. Figure \ref{Fig:FiftheOrderEffect}(a) shows the fluxes from numerical simulations for different frequency ratios and amplitudes, and these results verify pumping is fifth order in amplitude for frequency ratios $3/2$ and $4$.  Moreover, the fluxes for frequency ratios $3/2$ and $4$ are significantly smaller than the fluxes for frequency ratio $2$. For example, at amplitude $0.7$, the flux of frequency ratio $2$ is approximately $41$ times larger than the flux at frequency ratio $3/2$. In Figure \ref{Fig:FiftheOrderEffect}(b) we show the passive tracer particles over $10$ periods, and the net flow is not obvious without zooming in closely or playing movie 3 from the supplementary materials. Though the fluxes for frequency ratios $3/2$ and $4$ are small, they show clear scaling with amplitude. By contrast, the small fluxes for frequency ratio $3$ are not correlated with the amplitude. According to Result \ref{NonAntiperiodicTheorem}, there is no pumping for frequency ratio $3$, and the small fluxes reported in Figure \ref{Fig:FiftheOrderEffect}(a) arise from numerical error. 
Note that in Section \ref{sec:Section2.2} we examined how the flux scaled with amplitude for $\alpha=2$ at Re = 10. Here, we use Re = 40 because at lower Reynolds numbers, the fluxes are small, and it is difficult to measure pumping at higher order.

%
%
\subsection{Near\--Miss Frequencies}

\begin{figure}
    \centering
    \includegraphics[width=0.9\linewidth]{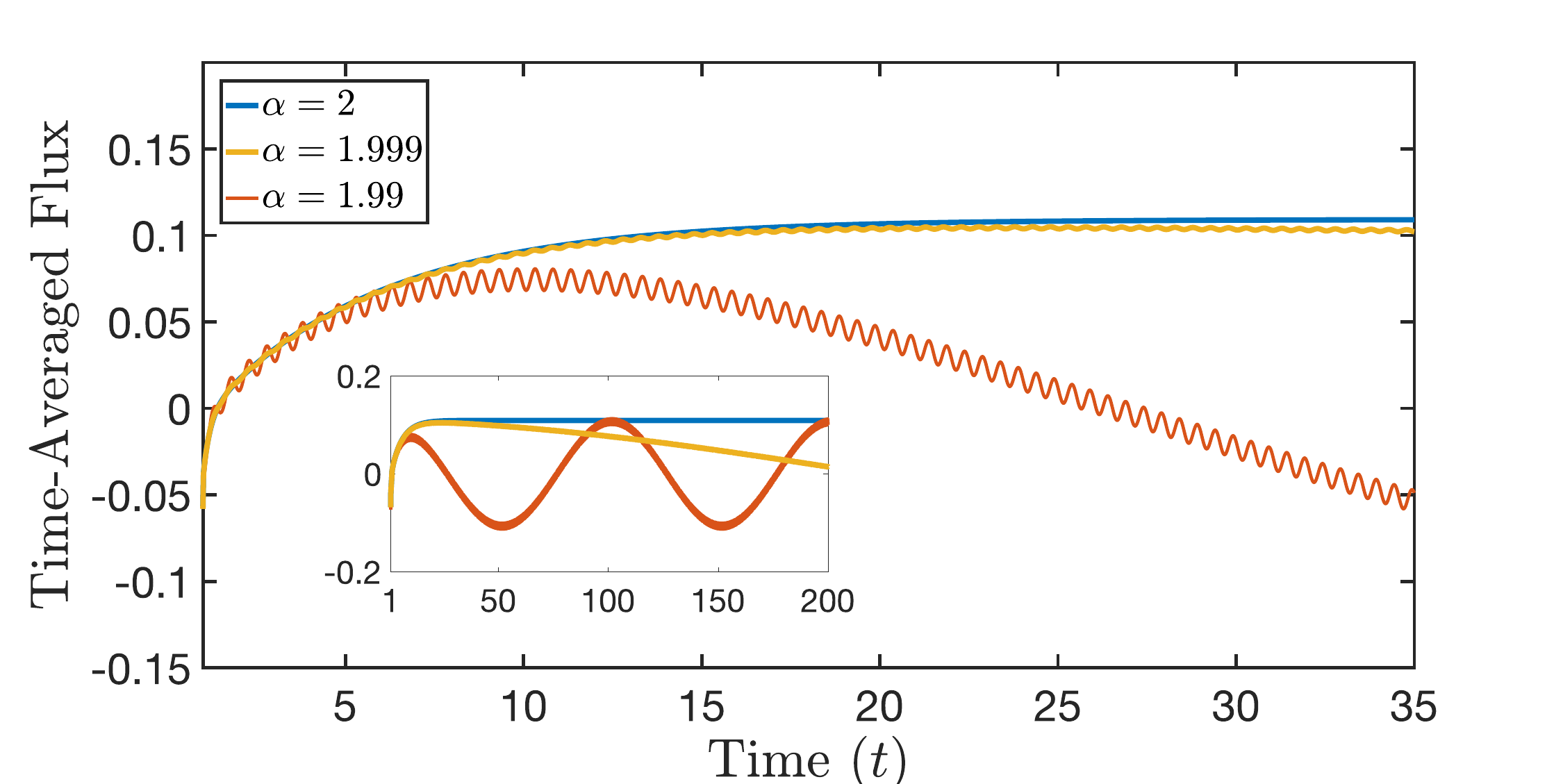}
    \caption{Time average of fluxes over one time unit for frequency ratios $\alpha$ = 2, $\alpha$ = 1.999, and $\alpha$ = 1.99 at amplitude $\epsilon = 0.7$, and Re = 10. 
    The inset shows the time\--averaged fluxes to time $200$.
    On short time scales the fluxes are similar, but on long time scales, the fluxes for $\alpha = 1.99$ and $\alpha = 1.999$ oscillate with amplitude similar to the steady flux for $\alpha = 2$.}
    \label{Fig:NearMiss}
\end{figure}

An interesting question is what happens for frequency ratios that are close to, but not equal to, $\alpha = 2$.
Figure \ref{Fig:NearMiss} shows the fluxes averaged over one time unit for frequency ratios $\alpha$ = 2, $\alpha$ = 1.999, and $\alpha$ = 1.99 at amplitude $\epsilon = 0.7$, and Re = 10. The fluxes for $\alpha = 1.99$ and $\alpha = 1.999$ oscillate on long time scales, with periods 100 and 1000, respectively, with amplitude similar to the steady flux for $\alpha = 2$. According to Result \ref{NonAntiperiodicTheorem}, for $\alpha = 1.99$ and $\alpha = 1.999$ the lowest order in amplitude pumping could occur is 299 and 2999, respectively. Over long periods we expect negligible net flux for these ``near\--miss'' frequency ratios, but on a short time scale they exhibit fluxes similar to $\alpha = 2$.

\section{Discussion}
Using asymptotic analysis and numerical simulations, we examined the flow around a cylinder moving with two-frequency collinear oscillations, and we derived necessary conditions for pumping. 
Strikingly, the necessary conditions on the frequency ratio for non-zero net motion are identical to those observed in the physical system of objects sliding via frictional forces on a laterally vibrating surface \citep{reznik2001c,hashemi2022net}.
In \cite{hashemi2022net}, they identified the time asymmetry of the driving oscillation necessary for net motion. Specifically, the sliding object only exhibited net motion when the vibrations were \textit{non-antiperiodic}. A $2\pi$-periodic function $f(t)$ is \textit{antiperiodic} if there exists $0<\phi<2\pi$ such that 
\begin{equation}
    f(t) = -f(t + \phi).
    \label{eq:defOfAntiperiodic}
\end{equation}
Conversely, $f(t)$ is non-antiperiodic if $f(t)$ does not satisfy Equation \eqref{eq:defOfAntiperiodic} for any $\phi$. \cite{hashemi2022net} showed that for frequency ratio $\alpha = b/a$, the two-frequency oscillation in equation \eqref{Eq:PositionOfCylinder} is antiperiodic if both $a$ and $b$ are odd. Otherwise, the motion is non-antiperiodic; i.e.\ one of $a$ or $b$ is odd and the other is even.  For example, for $\alpha = 1/1$ and $\alpha = 3/1$, the motion is antiperiodic (Equation \eqref{eq:defOfAntiperiodic} is satisfied for $\phi=\pi$), and for $\alpha=2/1$ the motion is non-antiperiodic.  

\begin{figure}
    \centering
    \includegraphics[width=1\linewidth]{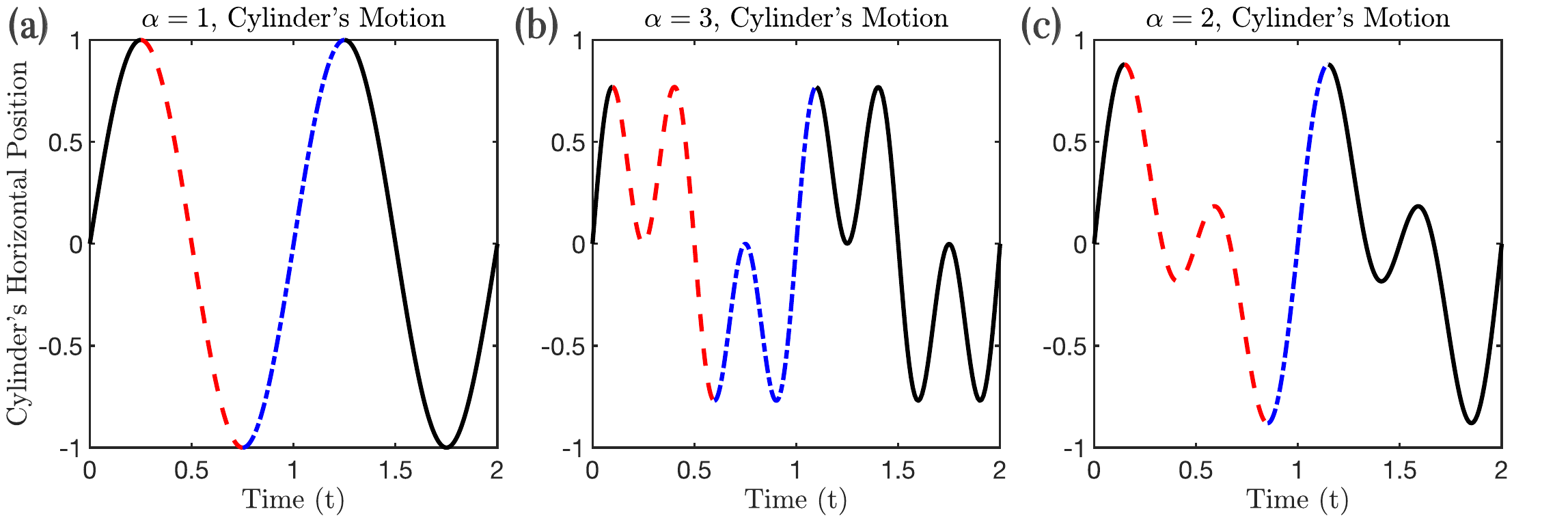}
    \caption{The displacement of the cylinder's center, defined by Equation \eqref{Eq:PositionOfCylinder}, is plotted from $0 \leq t \leq 2$ for (a) $\alpha = 1$, (b) $\alpha = 3$, and (c) $\alpha = 2$ when $A = 1$. The blue dashed dotted-curve corresponds to the path from the left-most to the right-most position of the cylinder, and the red dashed-curve shows the path from the right-most to the left-most position. }
    \label{Fig:IntuitiveObservation}
\end{figure}

To help illustrate the significance of antiperiodicity, in Figure \ref{Fig:IntuitiveObservation} we plot the cylinder's horizontal position for $\alpha = 1/1$, $\alpha = 3/1$, and $\alpha = 2/1$ and highlight the paths between the maximum and minimum position. For an antiperiodic oscillation (e.g.\ $\alpha=1$ and $\alpha=3$), there is a symmetry between the path from the right-most to the left-most position (red dashed-line) and the path from the left-most to right-most position (blue dashed dotted-line). Based on this symmetry, one would not expect a net flow. 
For a non-antiperiodic oscillation (e.g.\ $\alpha=2$), the cylinder’s motion from the right-most to left-most position is distinct from its counterpart. In our simulations, we observed this motion produced net flow to the right. 
One may suspect that the direction of the flow results from the larger average velocity along the path from its left-most to right-most position compared to its opposite. This idea is consistent with the observed net flow to the left when $\alpha = -2$. The relationship between the waveform and the direction of flow is likely not so simple, and further analysis is required. 
  
In this paper, we not only derived the necessary conditions for pumping but also obtained the smallest order in amplitude at which pumping occurs. Frequency ratio two results in pumping at third order in amplitude, and all other frequency ratios produce pumping at higher order. This result indicates that pumping is strongest for frequency ratio two at small amplitude. In numerical simulations, we observed (Figure \ref{Fig:FiftheOrderEffect}(a)) that the flux for $\alpha = 2$ is at least $10$ times larger than the flux from other frequency ratios.  These results are consistent with those of \cite{hashemi2022net} who reported both experimentally and numerically that $\alpha = 2$ generated the largest net motion. Likewise, \cite{reznik2001c} and \cite{hui2024vibrational} showed in their models that the largest net frictional force was produced when the frequency ratio was $\alpha = 2$.

Some insight into the connection between the flow around an oscillating cylinder and the frictional sliding problem is provided by \cite{hashemi2024multiple} in which they analyze the movement of a particle in a tank of fluid undergoing two\--frequency oscillations.  The model involves linear and nonlinear drag forces induced by the relative motion of the particle and the surrounding fluid. They perform analysis in the small amplitude limit and show that only frequency ratio $\alpha=2$ produces a nonzero time\--average velocity at first\--order in amplitude. Like the streaming problem we analyze, other frequency ratios produce net movement at higher order in amplitude.  \cite{hashemi2024multiple} argues that both models of particle in an oscillating tank and of an object sliding on frictional surface are related to the same reduced model of a forced mechanical system with nonlinear dissipation.



In our analysis and computational experiments, the two oscillations were of equal amplitude, in phase, and collinear.
Analysis and experiments on sliding frictional systems have considered the more general motion of the form 
\begin{equation}
    X(t) = \frac{A_1}{2}\sin(\Omega_1 t) + \frac{A_2}{2}\sin(\Omega_2 t + \phi).
    \label{Eq:MoreGeneralOscillatoryMotion}
\end{equation}
For example, it was predicted that for frequency ratio two, the maximum sliding velocity occurs for $A_2/A_1 \approx 0.25$
\citep{zhang2024theoretical,reznik1998coulomb}, which was validated experimentally \citep{hui2024vibrational}. The analysis from this work focused on necessary conditions on the frequency ratio for pumping based on the solution structure, and it does not predict how the pumping rate depends on parameters. The analytic solution can be obtained through second order \citep{holtsmark1954boundary}, but solving the resulting equations at third order and higher, where pumping occurs, is not feasible.
The pumping rate for different amplitudes, phases, and Reynolds numbers could be measured experimentally or numerically. 
We note that to our knowledge there have been no experiments which examine the pumping rate generated by a cylinder oscillating with two frequencies. 
Another interesting generalization of the motion involves non-collinear, two-frequency vibrations (i.e., oscillations with different frequencies in two different directions). It may be possible to extend the analysis of this work to derive necessary conditions for a temporal asymmetry, and hence a net\--directed flow, but more work on this problem is needed.

Steady streaming has been exploited in many microfluidic applications such as drug delivery \citep{sumner2021steady}, particle trapping \citep{agarwal2018inertial, mutlu2018oscillatory, patel2014cavity, thameem2016particle, volk2020size, wang2012efficient, zhang2024particle}, 
bubble-driven flow \citep{marmottant2006microfluidics, rallabandi2014two, wang2013frequency}, 
mixing \citep{ahmed2009fast, huang2013acoustofluidic, kumar2011steady, liu2003acoustic}, and 
pumping \citep{huang2014reliable, marmottant2004bubble, tovar2009lateral, tovar2011lateral, zhang2024elasto}.
Microchannel pumps driven by single-frequency steady streaming rely on design asymmetry of the apparatus to produce a net-directed flow. For example, \cite{tovar2009lateral} designed a micropump in which cavities of trapped air bubbles were placed at an angle along the channel walls. The angling of the cavities produces an asymmetric streaming flow that results in a net force that drive the flow through the channel. In our work, we have shown that a cylinder vibrating with two-frequency oscillations can pump fluid in either direction. The asymmetry in this case is temporal rather than spatial, and this idea could lead to new designs of microfluidic pumps or potentially enhance the performance of existing designs.

The autonomous propulsion of microparticles is a means of transporting objects by steady streaming \citep{ahmed2016density,collis2017autonomous,li2024propulsion,lippera2019no,nadal2014asymmetric,nadal2020acoustic,sabrina2018shape,wang2012autonomous}. The particle or the background flow vibrates with single-frequency oscillations, and variation in particle shape or density leads to an asymmetric streaming flow that induces a net propulsive force on the particle. The asymptotic analysis of this phenomenon involves computing how the particle asymmetry leads to a net force on itself \citep{nadal2014asymmetric,collis2017autonomous}.  Our analysis illustrates how a net force can arise from temporal asymmetry through two-frequency oscillations of a symmetric object. Though our analysis was for a cylinder, it could be extended to a sphere.  Based on our results, we predict that symmetric particles could be transported by employing two-frequency oscillations.

\vspace{1cm}
\noindent\textbf{Funding:} The work of W.D.R.\ was partially supported by the National Science Foundation under Grant No.\ CBET\-2125806.

\vspace{0.3cm}

\noindent\textbf{Declaration of Interest:} The authors report no conflict of interest.

\vspace{0.3cm}

\noindent\textbf{Data Availability Statement:} The source code of the numerical simulations is available from the corresponding author, R.D.G. The movies are available at \url{https://app.box.com/s/3guw4y6zpqdg9yyqrqqsrdkj1oy22cq8}.

\appendix
\section{Spatial Structure of the Solution} \label{sec:SectionA}
We prove Result \ref{thm:FirstTheorem} in Section \ref{sec:Section5} by proving a more general case involving multifrequency oscillations.  Suppose the cylinder is oscillating in the horizontal direction,  
$$X(t) = \epsilon\sum_{k=1}^{K} a_k\sin(kt + \phi_k).$$
Then fixing the cylinder, the governing equations for multifrequency oscillation are 
\begin{gather*}
    \left(\Delta - Re\frac{\partial}{\partial t}\right)\Delta \psi = \epsilon Re\;\boldsymbol{u}\bcdot\bnabla\left(\Delta\psi\right),\\
    \psi(r = 1) = 0, \quad \frac{\partial\psi}{\partial r}(r = 1) = 0,\\
    \psi \sim \epsilon\sum_{k = -K}^{K}c_k r e^{ikt}\sin(\theta),
\end{gather*}
where 
\begin{equation*}
c_k = \frac{-sgn(k)a_k k e^{isgn(k)\phi_k}}{2}, \quad c_0 = 0.
\end{equation*}
Expanding in the small amplitude limit, 
\begin{equation*}
\psi = \epsilon\psi_1 + \epsilon^2\psi_2 + \epsilon^3\psi_3 + O\left(\epsilon^4\right),
\end{equation*}
the equations at the $n^{th}$ order are
\begin{equation}
    \Delta\left(\Delta - Re\;\frac{\partial}{\partial t}\right) \psi_n = \sum_{i+j = n} Re\;\boldsymbol{u_i}\bcdot\bnabla(\Delta \psi_j),
    \label{Eq:GeneralPerturbation1}
\end{equation}
\begin{equation}
    \psi_n(r = 1) = 0, \quad \frac{\partial\psi_n}{\partial r}(r = 1) = 0,
    \label{Eq:GeneralPerturbation2}
\end{equation}
\begin{equation}
    \label{Eq:GeneralPerturbation3}
    \psi_n \sim
        \begin{cases}
            \sum_{k = -K}^{K}\left(c_k r e^{ikt}\right)\sin(\theta) & n = 1\\
            \\
            0 & n > 1.
        \end{cases}
\end{equation}
From the stream function, the velocity is computed from \eqref{eq:StreamFunctionRep}, and the pressure can be found from substituting the velocity into the momentum equation \eqref{Eq:NonDimMomentumEquation}.

To prove Result \ref{thm:FirstTheorem}, we prove two lemmas in the following subsections that show that terms proportional to $\sin(\theta)$ and $\cos(\theta)$ only appear at odd order in the expansions for the velocity and pressure, respectively.  In Section \ref{sec:Section4.4} we showed that a net force requires such terms, which together with the below lemmas prove Result \ref{thm:FirstTheorem}.

\subsection{Solution Structure of $\psi_n$}
Equation \eqref{Eq:GeneralPerturbation1} has the general solution, 
\begin{equation}
    \psi_n = \sum_{k,m=-\infty}^\infty f^{n}_{k,m}(r)e^{ikt}e^{im\theta},
    \label{Eq:GeneralPerturbationSolution}
\end{equation}
but the boundary conditions limit the solution to a finite number of terms. 
For example, the far-field boundary condition \eqref{Eq:GeneralPerturbation3} dictates the structure of $\psi_1$,
\begin{equation*}
\psi_1 = \left(\sum_{k = -K}^{K}f_{k,1}^1(r)e^{ikt}\right)\sin(\theta) = F_1^1(r, t)\sin(\theta).
\end{equation*}
Furthermore, the solution of $\psi_2$ is dependent on the quadratic nonlinearity of $\psi_1$ with itself, resulting in
\begin{equation}
\psi_2 = \left(\sum_{k = -2K}^{2K}f_{k,2}^2(r)e^{ikt}\right)\sin(2\theta) = F_2^2(r, t)\sin(2\theta).
\end{equation}
Generally, the interaction of lower order terms determines the structure of $\psi_n$, and the following Lemma shows that the solution structure depends on the parity of $n$. 

\begin{lemma}
\label{Thm:LocationOfSine}
For $n \in \mathbb{N}$, the solution of \eqref{Eq:GeneralPerturbation1} with its boundary conditions \eqref{Eq:GeneralPerturbation2} and \eqref{Eq:GeneralPerturbation3} has the form,
\begin{equation}
        \label{Eq:OddEvenSolution}
    \psi_n \sim
        \begin{cases}
            \sum_{l=1}^{n/2}F_{2l}^n (r,t)\sin(2l\theta) & \textbf{for n even} \\
            \\
            \sum_{l=0}^{(n-1)/2}F_{2l+1}^n (r,t)\sin\left((2l+1)\theta\right) & \textbf{for n odd}.
        \end{cases}
\end{equation}
\end{lemma}
\textbf{Proof:}
    We will proceed with induction, and choose $n = 1$ and $n = 2$ as the base cases.
    From the boundary conditions \eqref{Eq:GeneralPerturbation2} and \eqref{Eq:GeneralPerturbation3},
    \begin{equation*}
    \psi_1 = F_1^1(r,t)\sin(\theta).
    \end{equation*}
    To solve for $\psi_2$, we compute the right-hand side of \eqref{Eq:GeneralPerturbation1},
    \begin{equation*}
    Re\;\boldsymbol{u_1}\bcdot\bnabla\left(\Delta\psi_1\right) = \frac{Re}{2}\left(\frac{1}{r}F_1^1(r,t) \cdot \frac{\partial}{\partial r}D_1F_1^1(r, t) - \frac{1}{r}\frac{\partial}{\partial r}F_1^1(r,t) \cdot D_1 F_1^1(r, t)\right)\sin(2\theta),
    \end{equation*}
    where 
    \begin{equation*}
    D_1 = \frac{\partial^2}{\partial r^2} + \frac{1}{r}\frac{\partial}{\partial r} - \frac{1}{r^2}.
    \end{equation*}
    Therefore,
    \begin{equation*}
    \psi_2 = F_{2}^2(r,t)\sin(2\theta).
    \end{equation*}
    By induction, we assume \eqref{Eq:OddEvenSolution} holds true up to $\psi_{n-1}$, and, 
    for simplicity, assume $n$ is odd. Proving the even case involves the same procedure. 
    Note, each term of the right-hand side of (\ref{Eq:GeneralPerturbation1}) consists of
    \begin{equation}
        \label{eq:rightHandSideDerivatives}
        \boldsymbol{u_i}\bcdot\bnabla\left(\Delta\psi_j\right) = \frac{1}{r}\frac{\partial \psi_i}{\partial \theta}\frac{\partial}{\partial r}\left(\Delta\psi_j\right) - \frac{1}{r}\frac{\partial \psi_i}{\partial r}\frac{\partial}{\partial \theta}\left(\Delta\psi_j\right),
    \end{equation}
    where $i + j = n$.
    If we show (\ref{eq:rightHandSideDerivatives}) is a linear combination of $\{\sin(\theta), \sin(3\theta), \cdots, \sin(n\theta)\}$, then we have completed the induction. 

    Because $n$ is odd, one of $i$ and $j$ is even and the other is odd.  Without a loss of generality, choose $i$ even and $j$ odd. 
    Substituting $\psi_i$ and $\psi_j$ into  \eqref{eq:rightHandSideDerivatives}, we obtain 
\begin{equation}
\begin{array}{ll}
\boldsymbol{u_i}\bcdot\bnabla\left(\Delta\psi_j\right) &= \displaystyle\sum_{s=1}^{i/2}\sum_{q=0}^{(j-1)/2}\frac{2s}{r}F_{2s}^i(r,t)\frac{\partial}{\partial r}D_{2q+1}F_{2q+1}^j(r,t)\sin((2q+1)\theta)\cos(2s\theta)\\ [10pt]
\displaystyle &- \sum_{s=1}^{i/2}\sum_{q=0}^{(j-1)/2}\frac{(2q+1)}{r}D_{2q+1}F_{2q+1}^j(r,t)\frac{\partial}{\partial r}F_{2s}^i(r,t)\sin(2s\theta)\cos((2q+1)\theta),
 \end{array}
\label{eq:PlugIntoQuadNonLin}
\end{equation}
where 
\begin{equation*}
D_m = \frac{\partial^2 }{\partial r^2} + \frac{1}{r}\frac{\partial}{\partial r} - \frac{m^2}{r^2}.
\end{equation*}
We use the identity,
    \begin{equation}
        2\sin(a\theta)\cos(b\theta) = \sin((a+b)\theta) + \sin((a-b)\theta), \quad a, b \in \mathbb{N}
        \label{eq:trigIdentity}
    \end{equation}
    to simplify Equation (\ref{eq:PlugIntoQuadNonLin}) to 
    \begin{equation}
        \boldsymbol{u_i}\bcdot\bnabla\left(\Delta\psi_j\right) = \sum_{s = 1}^{i/2}\sum_{q=0}^{(j-1)/2} A_{s,q}^{i,j^-}(r,t)\sin((2q+2s+1)\theta) + A_{s,q}^{i,j^+}(r,t)\sin((2q-2s+1)\theta),
        \label{eq:PlugIntoQuadNonLin2}
    \end{equation}  
    where
    \begin{equation*}
    A_{s,q}^{i,j^\pm}(r,t) = \frac{s}{r}F_{2s}^i(r,t)\frac{\partial}{\partial r}D_{2q+1}F_{2q+1}^j(r,t) \pm \frac{(2q+1)}{2r}D_{2q+1}F_{2q+1}^j(r,t)\frac{\partial}{\partial r}F_{2s}^i(r,t).
    \end{equation*}
    For $1\leq s \leq i/2$ and $0 \leq q \leq (j-1)/2$,
    \begin{equation*}
    3 \leq |2q+2s+1|\leq i + j = n, \quad 1 \leq |2q-2s+1| \leq \max\{|i-3|, |j-2|\}.
    \end{equation*}
    In particular, 
    \begin{equation*}
    1 \leq |2q+2s+1|, \;|2q-2s+1|\leq n.
    \end{equation*}
    Therefore, for some coefficients $F_{2l+1}^n (r,t)$,
    \begin{equation*}
    \boldsymbol{u_i}\bcdot\bnabla\left(\Delta\psi_j\right) = \sum_{l=0}^{(n-1)/2}F_{2l+1}^n (r,t)\sin\left((2l+1)\theta\right),
    \end{equation*}
    which completes the induction.

\subsection{Solution Structure of $p_n$}
To compute the pressure, we convert the stream function, $\psi = \epsilon\psi_1 + \epsilon^2\psi_2 + \epsilon^3\psi_3 + \cdots$, into velocity components, 
\begin{subeqnarray}
u_r &=& \epsilon u_{r, 1} + \epsilon^2 u_{r, 2} + \epsilon^3 u_{r, 3} + O\left(\epsilon^4\right),\\[3pt]
u_\theta &=& \epsilon u_{\theta, 1} + \epsilon^2 u_{\theta, 2} + \epsilon^3 u_{\theta, 3} + O\left(\epsilon^4\right),
\label{Eq:VelocityComponentExpansion}
\end{subeqnarray}
where
\begin{equation}
u_{r, i} = \frac{1}{r}\frac{\partial \psi_i}{\partial \theta}, \quad u_{\theta, i} = -\frac{\partial \psi_i}{\partial r}. \nonumber
\end{equation}
We expand the pressure in the small amplitude limit, 
\begin{equation}
p = \epsilon p_1 + \epsilon^2 p_2 + \epsilon^3 p_3 + O\left(\epsilon^4\right),
\label{Eq:pressureExp}
\end{equation}
and substitute the velocity components \eqref{Eq:VelocityComponentExpansion} and the pressure \eqref{Eq:pressureExp} into the momentum equation \eqref{Eq:NonDimMomentumEquation} to obtain successive equations at each order $n$, 
\begin{subeqnarray}
\frac{\partial p_n}{\partial r} &=& \Delta u_{r, n} - \frac{u_{r, n}}{r^2} - \frac{2}{r^2}\frac{\partial u_{\theta, n}}{\partial \theta} - Re\frac{\partial u_{r, n}}{\partial t} - Re\sum_{i+j = n}\left(u_{r, i}\frac{\partial u_{r, j}}{\partial r} + \frac{u_{\theta, i}}{r}\frac{\partial u_{r, j}}{\partial \theta} - \frac{u_{\theta, i}u_{\theta, j}}{r}\right),\\[7pt]
\frac{1}{r}\frac{\partial p_n}{\partial \theta} &=& \Delta u_{\theta, n} - \frac{u_{\theta, n}}{r^2} + \frac{2}{r^2}\frac{\partial u_{r, n}}{\partial \theta} - Re \frac{\partial u_{\theta, n}}{\partial t} - Re\sum_{i+j=n}\left(u_{r, i}\frac{\partial u_{\theta, j}}{\partial r} + \frac{u_{\theta, i}}{r}\frac{\partial u_{\theta, j}}{\partial \theta} + \frac{u_{r, i}u_{\theta, j}}{r}\right).
\label{Eq:CylindricalNavier}
\end{subeqnarray}
For future computations, it is convenient to work with the linear and nonlinear terms separately so that 
\begin{subeqnarray}
\frac{\partial p_n}{\partial r} &=& L_{r, n} + Re \sum_{i+j=n}NL_{r, i, j},\\
\frac{1}{r}\frac{\partial p_n}{\partial \theta} &=& L_{\theta, n} + Re \sum_{i+j=n}NL_{\theta, i, j},
\label{Eq:CylindricalNavier2}
\end{subeqnarray}
where
\begin{subequations}
\begin{align}
    L_{r, n} &=  \Delta u_{r, n} - \frac{u_{r, n}}{r^2} - \frac{2}{r^2}\frac{\partial u_{\theta, n}}{\partial \theta} - Re\frac{\partial u_{r, n}}{\partial t}, \\
L_{\theta, n} &= \Delta u_{\theta, n} - \frac{u_{\theta, n}}{r^2} + \frac{2}{r^2}\frac{\partial u_{r, n}}{\partial \theta} - Re \frac{\partial u_{\theta, n}}{\partial t} \\
NL_{r, i, j} &= u_{r, i}\frac{\partial u_{r, j}}{\partial r} + \frac{u_{\theta, i}}{r}\frac{\partial u_{r, j}}{\partial \theta} - \frac{u_{\theta, i}u_{\theta, j}}{r}, \label{Eq:rCompNL} \\
NL_{\theta, i, j} &= u_{r, i}\frac{\partial u_{\theta, j}}{\partial r} + \frac{u_{\theta, i}}{r}\frac{\partial u_{\theta, j}}{\partial \theta} + \frac{u_{r, i}u_{\theta, j}}{r}. \label{Eq:tCompNL}
\end{align}
\end{subequations}

The structure of the expansion of the pressure is given by the below lemma. 
\begin{lemma}
\label{Lemma2}
For $n \in \mathbb{N}$, the pressure from \eqref{Eq:CylindricalNavier} has the form,
\begin{equation}
    \label{Eq:PressureSolutionStructure}
    p_n =
        \begin{cases}
            \sum_{l=0}^{n/2}G_{2l}^n (r,t)\cos(2l\theta) & \textbf{for n even} \\
            \\
            \sum_{l=0}^{(n-1)/2}G_{2l+1}^n (r,t)\cos\left((2l+1)\theta\right) & \textbf{for n odd}.
        \end{cases}
\end{equation}
\end{lemma}
\textbf{Proof:}
We will examine the case when $n$ is odd, as showing the case when $n$ is even entails the same procedure. 
To prove the result, we show that $r$-component of \eqref{Eq:CylindricalNavier} can be expressed as a linear combination of $\{\cos(\theta), \cos(3\theta), \cdots, \cos(n\theta)\}$ and the $\theta$-component can be expressed as a linear combination of $\{\sin(\theta), \sin(3\theta), \cdots, \sin(n\theta)\}$. We show this is true first for the linear terms, and then for the nonlinear terms. 

We begin with the linear terms. From Lemma \eqref{Thm:LocationOfSine}, the stream function $\psi_n$ has the form,
\begin{equation}
\psi_n = \sum_{l=1}^{(n-1)/2}F_{2l+1}^n (r,t)\sin\left((2l+1)\theta\right).\nonumber
\end{equation}
Substituting $\psi_n$ into $L_{r, n}$, we obtain
\begin{equation}
L_{r, n} =  \sum_{l=1}^{(n-1)/2}M_{2l+1}^n(r,t)\cos\left((2l+1)\theta\right),
\label{Eq:LinearR}
\end{equation}
where
\begin{equation}
M_{m}^n(r,t) = -\frac{4m}{r^3}\left(-\frac{r^2}{4}\frac{\partial^2F_{m}^n}{\partial r^2} - \frac{r}{4}\frac{\partial F_{m}^n}{\partial r} + \frac{m^2}{4} F_{m}^n+ \frac{Re\; r^2}{4}\frac{\partial F_{m}^n}{\partial t}\right).\nonumber
\end{equation}
Similarly for $L_{\theta, n}$,
\begin{equation}
L_{\theta,n} =  \sum_{l=1}^{(n-1)/2}S_{2l+1}^n(r,t)\sin\left((2l+1)\theta\right),
\label{Eq:LinearTheta}
\end{equation}
where 
\begin{equation}
S_{s}^n(r,t) = -\frac{\partial^3}{\partial r^3}F_s^n - \frac{1}{r}\frac{\partial^2F_s^n}{\partial r^2} + \left(\frac{s}{r}\right)^2\frac{\partial F_s^n}{\partial r} + \frac{1}{r^2}\frac{\partial F_s^n}{\partial r}-\frac{2s^2}{r^3}F_s^n + Re\frac{\partial^2F_s^n}{\partial t \partial r}.\nonumber
\end{equation}

We next examine the nonlinear terms $NL_{r, i, j}$ and $NL_{\theta, i, j}$ for $i+j=n$. Because $n$ is odd, one of $i$ or $j$ is even and the other is odd. Without a loss of generality, choose $i$ even and $j$ odd.  Using Lemma \eqref{Thm:LocationOfSine} for the form of the stream function, we substitute the velocity into Equation \eqref{Eq:rCompNL} for $NL_{r, i, j}$ to obtain
\begin{equation}
NL_{r, i, j}=\sum_{s = 1}^{i/2}\sum_{q = 0}^{(j-1)/2}A_{i, j}(r, t)\sin(2s\theta)\sin((2q+1)\theta) + B_{i, j}(r, t)\cos(2s\theta)\cos((2q+1)\theta),
\label{Eq:rCompNL2}
\end{equation}
where 
\begin{equation}
A_{i, j}(r, t) = -\frac{1}{r}\frac{\partial}{\partial r}F_{2s}^i(r, t)\frac{\partial}{\partial r}F_{2q+1}^j(r, t) + \frac{(2q+1)^2}{r^2}\frac{\partial }{\partial r}F_{2s}^i(r, t)F_{2q+1}^j(r, t), \nonumber
\end{equation}
and
\begin{equation}
B_{i, j}(r, t) = \frac{2s}{r^3}(2q+1)F_{2s}^i(r, t)\left(r \frac{\partial}{\partial r}F_{2q+1}^j(r, t) - F_{2q+1}^j(r, t)\right).\nonumber
\end{equation}
For Equation \eqref{Eq:rCompNL2}, we invoke the following trigonometric identity. For $a, b \in \mathbb{N}$,
\begin{subeqnarray}
2\sin(a\theta)\sin(b\theta) &=& \cos((a-b)\theta) - \cos((a+b)\theta),\\[3pt]
2\cos(a\theta)\cos(b\theta) &=& \cos((a-b)\theta) + \cos((a+b)\theta).
\label{eq:trigIdentity2}
\end{subeqnarray}
Because $1 \leq s \leq i/2$ and $1 \leq q \leq (j-1)/2$, 
\begin{equation}
3 \leq \left|2q + 2s + 1\right|\leq i + j = n, \quad 1 \leq \left|2q - 2s + 1\right| \leq \max\{\left|i-3\right|, \left|j-2\right|\},\nonumber
\end{equation}
and Equation \eqref{Eq:rCompNL2} is equivalent to 
\begin{equation}
NL_{r, i, j}=\sum_{l = 0}^{(n-1)/2}J_{2l+1}^n(r, t)\cos((2l+1)\theta),
\label{Eq:rCompNL3}
\end{equation}
for some functions $J_{2l+1}^n(r, t)$.

Similarly, we substitute the velocity into $NL_{\theta, i, j}$ of Equation \eqref{Eq:tCompNL} to obtain 
\begin{equation}
NL_{\theta, i, j}=\sum_{s = 1}^{i/2}\sum_{q = 0}^{(j-1)/2} E_{i, j}(r,t)\sin((2q+1)\theta)\cos(2s\theta) + F_{i, j}(r, t)\sin(2s\theta)\cos((2q+1)\theta),
\label{Eq:tCompNL2}
\end{equation}
where 
\begin{equation}
E_{i, j}(r, t) = -\frac{2s}{r}F_{2s}^i(r, t)\frac{\partial^2}{\partial r^2}F_{2q+1}^j(r, t)-\frac{2s}{r^2}F_{2s}^i(r, t)\frac{\partial}{\partial r}F_{2q+1}^j(r, t),\nonumber
\end{equation}
and
\begin{equation}
F_{i, j}(r, t) = \frac{2q+1}{r} \frac{\partial}{\partial r}F_{2s}^i(r, t)\frac{\partial}{\partial r}F_{2q+1}^j(r, t).\nonumber
\end{equation}
We simplify Equation \eqref{Eq:tCompNL2} using the trigonometric identity from Equation \eqref{eq:trigIdentity} to obtain
\begin{equation}
NL_{\theta, i, j}=\sum_{l = 0}^{(n-1)/2}K_{2l+1}^n(r, t)\sin((2l+1)\theta)
\label{Eq:tCompNL3}
\end{equation}
for some functions and $K_{2l+1}^n(r, t)$.

Finally, from \eqref{Eq:LinearR} and \eqref{Eq:rCompNL3}, $\partial p_n/\partial r$ is composed of cosine functions with odd arguments, and similarly, from \eqref{Eq:LinearTheta} and \eqref{Eq:tCompNL3}, $\partial p_n/\partial \theta$ is composed of sine functions with odd arguments. Integrating with respect to both $r$ and $\theta$, the solution structure of $p_n$ is
\begin{equation}
p_n = \sum_{l = 0}^{(n-1)/2}G_{2l+1}^n(r, t)\cos((2l+1)\theta).\nonumber
\end{equation}
Therefore, Equation \eqref{Eq:PressureSolutionStructure} has been proven.

\section{Establishing Necessary Conditions for Pumping}
In Section \ref{sec:Section4.4}, we showed that for frequency ratio $\alpha = 2$, the third order terms proportional to $\sin(\theta)$ in the steady, stream function (and $\cos(\theta)$ in the steady pressure) produce a net force. To extend the argument from Section \ref{sec:Section4.4} to general frequency ratios, we derive necessary conditions for the presence of a steady term proportional to $\sin(\theta)$ in the stream function or proportional to $\cos(\theta)$ in the pressure.  Lemmas \ref{Thm:LocationOfSine} and \ref{Lemma2} state that $\sin(\theta)$ and $\cos(\theta)$ terms can only appear when $n$ is odd.   \textit{Therefore, the necessary condition for pumping is equivalent to finding order $n$ so that $n$ is odd, and $\psi_n$ has a steady term.}

\subsection{Frequencies at each order}
We first show which frequencies occur at each order. Suppose the cylinder oscillates in the horizontal direction with the motion,  
\begin{equation}
    X(t) = \frac{\epsilon}{2}\left(\sin(at) + \sin(bt)\right),
    \label{eq:LinearCombFrequency1}
\end{equation}  
where $a, b, \in \mathbb{Z}$ and gcd$(a, b) = 1$. Examining the first order solution \eqref{Eq:TwoFrequencyFirstOrderSolution}, $\psi_1$ involves the terms proportional to $e^{\pm iat}$ and $e^{\pm ibt}$. 
To solve for the second order solution \eqref{Eq:TwoFrequencySecondOrderSolution}, we substitute $\psi_1$ into the right-hand side of \eqref{Eq:NthOrderEquation}. From the products of the exponential functions, the frequencies arise at second order are all sums of $\pm a$ and $\pm b$. Generally, the solution structure of $\psi_n$ will be determined by the products of exponentials from lower order solutions. If $\psi_n$ contains a term proportional to $e^{ift}$, then $e^{ift}$ is related to $e^{\pm iat}$ and $e^{\pm ibt}$ by 
\begin{equation*}
e^{ift} = \left(e^{iat}\right)^{\xi_1}\left(e^{-iat}\right)^{\xi_2}\left(e^{ibt}\right)^{\xi_3}\left(e^{-ibt}\right)^{\xi_4},
\end{equation*}
and
\begin{equation}
    \xi_1 + \xi_2 + \xi_3 + \xi_4 = n, \quad \xi_i \geq 0, \quad i = 1, 2, 3, 4.
    \label{eq:LinearCombFrequency2}
\end{equation}
Therefore,
\begin{equation}
    f = \xi^{(1)}a + \xi^{(2)}b,
    \label{eq:LinearCombFrequency3}
\end{equation}
where 
\begin{align}
\xi^{(1)} = (\xi_1 - \xi_2), \label{freqA:eq} \\ 
\xi^{(2)} = (\xi_3 - \xi_4). \label{freqB:eq}
\end{align}

The above analysis is based on the structure of equation \eqref{Eq:NthOrderEquation} for the stream function. We next consider the pressure and show the pressure at $n^{th}$ order involves the same frequencies as the stream function.  To begin, we take the divergence of the Navier Stokes equation, and we use the divergence-free condition to obtain
\begin{equation}
\Delta p = -Re\; \bnabla \bcdot \left(\boldsymbol{u} \bcdot \bnabla \boldsymbol{u}\right).
\label{NSPressureForm}
\end{equation}
After expanding the solution in power of $\epsilon$, the pressure at $n^{th}$ order satisfies
\begin{equation}
\Delta p_n = -Re\sum_{i + j = n}\bnabla\bcdot\left(\boldsymbol{u_i}\bcdot \bnabla\boldsymbol{u_j}\right).
\label{eq:NSPressureForm2}
\end{equation}
The right side of this equation involves the same quadratic nonlinearity of the velocity that appears on the right side of 
equation \eqref{Eq:NthOrderEquation} for the stream function. Thus the stream function and the pressure involve the same frequencies at each order.

\subsection{Existence of Pumping}
We return to equations \eqref{eq:LinearCombFrequency2}\--\eqref{freqB:eq} which give the frequencies that occur at each order. A steady solution ($f=0$) occurs when 
\begin{equation*}
0 = \xi^{(1)}a + \xi^{(2)}b.
\end{equation*}
Solving this equation for $\xi^{(1)}$, we obtain 
\begin{equation*}
\xi^{(1)} = -\frac{b \xi^{(2)}}{a}.
\end{equation*}
Because gcd$(a, b) = 1$, and $\xi^{(1)}$ and $\xi^{(2)}$ are integers, solutions to this equation can be expressed as  
\begin{equation}
    \xi^{(1)} = \xi_{1}-\xi_{2} =cb, \qquad \xi^{(2)} =\xi_{3}-\xi_{4}= -ca, \qquad\qquad c\in\mathbb{Z}.
    \label{eq:LinearCombFrequency4}
\end{equation}  
Using these equations to eliminate $\xi_{1}$ and $\xi_{2}$ from \eqref{eq:LinearCombFrequency2} gives the order at which steady solutions occur as
\begin{equation}
n = 2\left(\xi_2 + \xi_3\right) + c(a + b).
\label{order_of_steady:eq}
\end{equation}
We use this last equation to prove Results \ref{AntiperiodicTheorem} and \ref{NonAntiperiodicTheorem}.

\subsubsection{Proof of Result \ref{AntiperiodicTheorem}}\label{sec:Antiperiodic}
Let $a$ and $b$ be both odd. From \eqref{order_of_steady:eq}, because $a+b$ is even, steady terms can only occur at even orders. From Result \ref{thm:FirstTheorem}, there is no pumping. 

\subsubsection{Proof of Result \ref{NonAntiperiodicTheorem} }\label{sec:Non-Antiperiodic} 
Without a loss of generality, choose $a$ odd and $b$ even. From \eqref{order_of_steady:eq}, steady terms occur at odd order $n$ for $c$ odd. The smallest odd order $n$ is obtained by choosing $c = 1$ and $\xi_2 = \xi_3 = 0$, which is order $n = a + b$.

 \label{sec:Pumping_proofs}
\section{Numerical Methods}\label{sec:NumericalMethods}
\subsection{Immersed Boundary Method}
We use the Immersed Boundary (IB) method to solve the Navier-Stokes equations. The IB method uses an Eulerian coordinate system for the fluid and a Lagrangian coordinate system for the immersed structures (i.e.\ cylinder or channel walls) \citep{peskin2002immersed}. Let $s$ be the parametric coordinate of a structure and $\boldsymbol{X}(s, t)$ be its position. We use capital letters $\boldsymbol{X}(s, t)$, $\boldsymbol{U}(s, t)$, and $\boldsymbol{F}(s, t)$ to define position, velocity, and force density in Lagrangian coordinates, and similarly we use lower case $p(\boldsymbol{x}, t)$, $\boldsymbol{u}(\boldsymbol{x}, t)$, and $\boldsymbol{f}(\boldsymbol{x}, t)$ for pressure, velocity, and force density in Eulerian coordinates.

The forces on the structures are applied to the surrounding fluid, and the fluid and structure move with the same velocity on the structure. The structural force density $\boldsymbol{F}(s, t)$ in Eulerian coordinates is given by
\begin{equation}
    \label{Eq:SpreadOperator}
    \boldsymbol{f}(\boldsymbol{x}, t) = S\boldsymbol{F} = \int_{\text{structure}}\boldsymbol{F}(s, t)\delta(\boldsymbol{x} - \boldsymbol{X}(s, t))\;ds,
\end{equation} 
where $\delta\left(\boldsymbol{x}\right)$ is the Dirac delta function. The operator $S$ ``spreads'' the the force density from the immersed structure to the surrounding fluid. Similarly the fluid velocity is interpolated to the immersed structure by the adjoint of the spreading operator:
\begin{align}
    \label{Eq:InterpolatingOperator}
    \boldsymbol{U}(s, t) = S^*\boldsymbol{u} = \int_{\text{fluid}}\boldsymbol{u}(\boldsymbol{x}, t)\delta(\boldsymbol{x} - \boldsymbol{X}(s, t))\;d\boldsymbol{x}.
\end{align}
In our simulations, the motion of the structure is prescribed, and the force density on the structure, $\boldsymbol{F}(s, t)$, is determined implicitly by requiring the fluid velocity match the prescribed velocity of the boundary: 
\begin{align*}
    S^*\boldsymbol{u} = \boldsymbol{U}_b.
\end{align*}
The full system describing the fluid and immersed boundaries is 
\begin{align}
    \label{Eq:MomentumIB}
    \rho\left(\frac{\partial \boldsymbol{u}}{\partial t} + \boldsymbol{u}\cdot \nabla \boldsymbol{u}\right) 
        &= -\nabla p + \mu \Delta \boldsymbol{u} + S\boldsymbol{F},\\
    \label{Eq:IncompressibilityIB}
    \nabla \cdot \boldsymbol{u} &= 0,\\
    \label{Eq:Contraint}
    S^*\boldsymbol{u} &= \boldsymbol{U}_b,
\end{align}
where $\rho$ is the fluid density and $\mu$ is the fluid viscosity.

\subsection{Discretization} \label{sec:Discretization}
We solve on a doubly periodic domain discretized into points equally spaced by $\Delta x$, and structures are discretized by points equally spaced by $\Delta s\approx\Delta x$.  We approximate the differential operators using Fourier pseudo-spectral methods. The discrete delta function is 
\begin{align*}
    \delta = \delta_{\Delta x}(x)\delta_{\Delta x}(y),
\end{align*}   
where
\begin{align}
    \label{Eq:DiscreteDelta}
    \delta_{\Delta x}(x)= \begin{cases} 
        \frac{1}{4\Delta x}\left(1 + \cos\left(\frac{\pi x}{2\Delta x}\right)\right) & |x| < 2\Delta x,\\
        \\
        0 & \text{else}. 
     \end{cases}
\end{align}
The discretized  spread (\ref{Eq:SpreadOperator}) and interpolating (\ref{Eq:InterpolatingOperator}) operators are
\begin{equation*}
    f_{i,j} = S\boldsymbol{F} = \Delta s\sum_{k}F_k\delta_{\Delta x}\left(x_i - X_k\right)\delta_{\Delta x}\left(y_j - Y_k\right),
\end{equation*}
\begin{equation*}
    U_k = S^*\boldsymbol{u} = \Delta x^2\sum_{i,j}u_{i, j}\delta_{\Delta x}\left(x_i - X_k\right)\delta_{\Delta x}(y_j - Y_k).
\end{equation*}

For the temporal discretization, we use a second-order IMEX scheme, named SBDF in \cite{ascher1995implicit}, in which the nonlinear terms are treated explilty in time, and the terms for the viscous force and structure force are treated implicitly in time with BDF2. The discretized system is 
\begin{align}
    \label{Eq:MomentumDiscrete}
    \rho\left(\frac{3\boldsymbol{u}^{n+1}-4\boldsymbol{u}^{n}+\boldsymbol{u}^{n-1}}{2\Delta t} + 2\boldsymbol{u}^n\cdot G\boldsymbol{u}^n - \boldsymbol{u}^{n-1}\cdot G\boldsymbol{u}^{n-1}\right) &= -Gp^{n+1} + \mu L\boldsymbol{u}^{n+1} + S\boldsymbol{F}^{n+1},\\
    \label{Eq:ContinuityDiscrete}
    D\boldsymbol{u}^{n+1} &= 0,\\
    \label{Eq:ConditionDiscrete}
    S^*\boldsymbol{u}^{n+1} &= \boldsymbol{U}_b^{n+1},
\end{align}
where $G, D,$ and $L$ represent the discrete gradient, divergence, and Laplacian, respectively. Because the nonlinear terms are treated explicitly in time, the resulting system to solve at each time step is linear. In block form, the system is
\begin{align}
    \label{Eq:MatrixForm}
    \begin{bmatrix}
        A & G & -S\\
        D & 0 & 0\\
        S^* & 0 & 0
    \end{bmatrix}
    \begin{bmatrix}
        \boldsymbol{u}\\
        p\\
        \boldsymbol{F}
    \end{bmatrix}^{n+1} = 
    \begin{bmatrix}
        \boldsymbol{g}\\
        0\\
        \boldsymbol{U}_b^{n+1}
    \end{bmatrix},
\end{align}
where 
\begin{equation*}
    A = \frac{3\rho}{2\Delta t}I - \mu L, 
\end{equation*}
and the known terms are
\begin{equation*}
    g = \rho\left(-2\boldsymbol{u}^n\cdot G\boldsymbol{u}^n + \boldsymbol{u}^{n-1}\cdot G\boldsymbol{u}^{n-1} - \frac{-4\boldsymbol{u}^n + \boldsymbol{u}^{n-1}}{2\Delta t}\right).
\end{equation*}

We solve \eqref{Eq:MatrixForm} at each time step by first solving for the force density which satisfies
\begin{equation}
    S^{*}\mathcal{L}^{-1}S\boldsymbol{F} =\boldsymbol{U}_b^{n+1} - S^*\mathcal{L}^{-1}\boldsymbol{g}, 
    \label{Eq:SC_for_force}
\end{equation}
where we denote the operator which maps the fluid force density to the fluid velocity by $\mathcal{L}^{-1}$. Specifically, suppose
$\boldsymbol{u}$ and $p$ satisfy the system
\begin{align*}
    \begin{bmatrix}
        A & G\\
        D & 0
    \end{bmatrix}
    \begin{bmatrix}
        \boldsymbol{u}\\
        p
    \end{bmatrix}
    +
    \begin{bmatrix}
        \boldsymbol{f}\\
        0
    \end{bmatrix}
    = 0,
\end{align*}
and $\mathcal{L}^{-1}$ is then 
\begin{equation*}
  \boldsymbol{u}=\mathcal{L}^{-1}\boldsymbol{f}.
\end{equation*}
The operator $\mathcal{L}^{-1}$ can be applied efficiently using the FFT. We solve equation \eqref{Eq:SC_for_force} with a Krylov method, which requires a preconditioner for efficiency. For preconditioning, we explicitly form the dense matrix representing $S^{*}\mathcal{L}^{-1}S$ at time $0$ and compute its Cholesky factorization. This work is performed before running the simulation. 

The grid resolution for channel simulations was $\Delta x=8/256$, so that there were $256$ grid points in the vertical direction, or equivalently $64$ points along the cylinder diameter. The time step was $\Delta{t}=5.00\times10^{-4}$. The tolerance for the conjugate gradient method used to solve \eqref{Eq:SC_for_force} was $1 \times10^{-3}$. 

\begin{table}
  \begin{center}
\def~{\hphantom{0}}
\setlength{\tabcolsep}{5pt}
  \begin{tabular}{lccccc}
        $\Delta x$  &  $\Delta t$  &  Flux  &  Difference  &  Ratio  &  $\log_2(\text{Ratio})$\\[5pt]
       8/256  &  $1.00\times10^{-2}$  &  $9.6044\times10^{-2}$ &---&  ---  & --- \\
       8/512  &  $5.00\times10^{-3}$  &  $1.0130\times10^{-1}$  &  $5.2573 \times 10^{-3}$  &  --- & --- \\
       8/1024  &  $2.50 \times10^{-3}$  &  $1.0279\times10^{-1}$  &  $1.4937 \times 10^{-3}$  &  3.5197  &  1.8155\\
       8/2048  &  $1.25 \times10^{-3}$  &  $1.0324\times10^{-1}$  &  $4.4149 \times 10^{-4}$  &  3.3833  &  1.7584\\
  \end{tabular}
  \caption{Refinement study for the flux with simultaneous refinement of space and time for the flow around an oscillating cylinder in a square 8 by 8 domain with periodic boundary conditions at $\text{Re} = 10$, amplitude $\epsilon = 0.5$, and time T = 100.} 
  \label{tab:Refine1}
  \end{center}
\end{table}

For the simulations presented in Figure \ref{fig:alpha=0_5}, the grid resolution was $\Delta x=8/512$ and the time step was $\Delta{t}=5.00\times10^{-3}$. Table \ref{tab:Refine1} shows the results of a mesh refinement study on this problem which shows that the time\--averaged flux converges at approximately second-order. Using the finest mesh $(\Delta x = 8/2048, \Delta t = 1.25 \times10^{-3})$ as the solution, we estimate the error in the flux for $\Delta x = 8/512$ and $\Delta t = 5.00\times10^{-3}$ is approximately $1.9\%$.

\begin{table}
  \begin{center}
\def~{\hphantom{0}}
\setlength{\tabcolsep}{5pt}
  \begin{tabular}{lcccc}
       $\Delta t$  &  Flux  &  Difference  &  Ratio  &  $\log_2(\text{Ratio})$\\[5pt]
       $1.000\times10^{-2}$  &  $9.6044\times10^{-2}$ &---&  ---  & --- \\
       $5.000\times10^{-3}$  &  $1.0086\times10^{-1}$  &  $4.8185 \times 10^{-3}$  &  --- & --- \\
       $2.500 \times10^{-3}$  &  $1.0175\times10^{-1}$  &  $8.8700 \times 10^{-4}$  &  5.4324  &  2.4416\\
       $1.250 \times10^{-3}$  &  $1.0191\times10^{-1}$  &  $1.6291 \times 10^{-4}$  &  5.4447  &  2.4448\\
       $6.250 \times10^{-4}$  &  $1.0194\times10^{-1}$  &  $3.2425 \times 10^{-5}$  &  5.0243  &  2.3289\\
       $3.125 \times10^{-4}$  &  $1.0195\times10^{-1}$  &  $7.1196 \times 10^{-6}$  &  4.5543  &  2.1872\\
  \end{tabular}
  \caption{Refinement study in time only for the flux with the space step fixed at $\Delta x = 8/256$ for the flow around an oscillating cylinder in a square 8 by 8 domain with periodic boundary conditions at $\text{Re} = 10$, amplitude $\epsilon = 0.5$, and time T = 100.} 
  \label{tab:Refine2}
  \end{center}
\end{table}

In the channel simulations, we used a coarser spatial resolution, $\Delta x = 8/256$, but a smaller time step $\Delta t = 5.00\times10^{-4}$.
Fixing the grid resolution as $\Delta x = 8/256$ and refining the time step, we find that the flux converges approximately second-order in $\Delta t$ (Table \ref{tab:Refine2}). For the smallest time step, the relative changes in flux are on the scale of $10^{-4}$, and thus the spatial error is dominant. Comparing with the fluxes from the space-time refinement study in  Table \ref{tab:Refine1}, we estimate the error in our numerical simulations to be on the scale of $1\--2\%$.

\bibliographystyle{Style}
\bibliography{myCite}

\end{document}